\documentclass{cernrep}
\usepackage{dcolumn}
\newcolumntype{d}{D{.}{.}{2.5}}
\newcolumntype{s}{D{.}{.}{1.2}}
\usepackage{varwidth}
\usepackage{xcolor}

\begin{document}

\title{Beam Transfer and Machine Protection}

\author{V. Kain}

\institute{CERN, Geneva, Switzerland}

\maketitle 

\begin{abstract}
Beam transfer, such as injection into or extraction from an accelerator, is one of the most critical moments in terms of machine protection in a high-intensity machine. Special equipment is used and machine protection aspects have to be taken into account in the design of the beam transfer concepts. A brief introduction of the principles of beam transfer and the equipment involved will be given in this lecture. The main concepts of machine protection for injection and extraction will be presented, with examples from the CERN SPS and LHC.
\end{abstract}

\section{Introduction}
\begin{figure}
\centering\includegraphics[width=.7\linewidth]{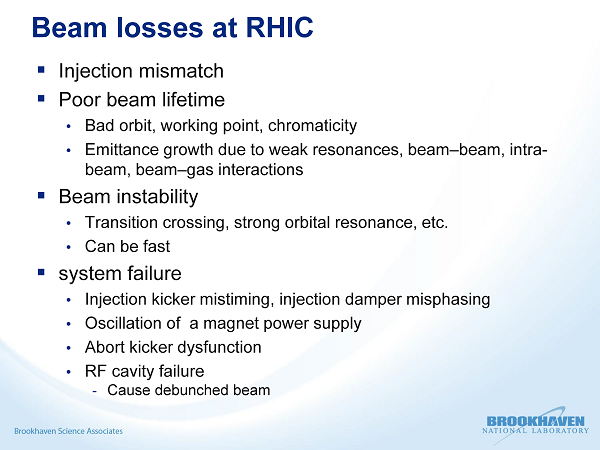}
\caption{Most frequent causes of beam loss at the Relativistic Heavy Ion Collider accelerator at Brookhaven National Laboratory. Issues during beam transfer, \eg injection and beam abort, are prominent. Presentation by M. Bei at the Workshop on Beam-induced Quenches at CERN \cite{bib:quench}.}
\label{fig:meiBei}
\end{figure}

Injection and extraction from an accelerator are critical operations on short time-scales with a high risk of significant beam loss in the case of equipment malfunction or badly adjusted parameters. Figure \ref{fig:meiBei}, which is taken from a presentation at an LHC workshop on beam-induced quenches, highlights how prominent losses induced by beam transfer actions are in the day-to-day operation of accelerators. The example is from the Relativistic Heavy Ion Collider accelerator at Brookhaven National Laboratory.
While in the first part of this lecture some of the typical injection, extraction, and transfer line configurations, along with associated equipment, are discussed, the second part is dedicated to a presentation of the machine protection systems developed for the LHC injection and beam dump system, as examples for machine protection systems for beam transfer.

\section{Injection}
Various different injection techniques exist in the different machines around the world. The chosen technique depends on the requirements in terms of loss acceptance, brightness, and particle type used. This section, as well as Sections~\ref{sec:Transfer lines} and \ref{sec:Extraction}, is based on the lectures by B.~Goddard, W.~Bartmann, M.~Barnes and M.~Meddahi at the CERN Accelerator School \cite{bib:cas}.

\subsubsection{Single-turn injection for hadrons}
This type of injection is typically used for transfer between machines in an accelerator chain in associ\-ation with \textit{boxcar stacking,} where a larger ring is filled sequentially by a smaller one. Angle and position errors at the injection point lead to injection oscillations, while optical errors lead to betatron mismatch. Both can cause emittance blow-up. As most of the machine protection concepts will be introduced in connection with this type of injection, it will be described in detail in Section~\ref{sec:Single-turn injection and injection equipment}.

\subsubsection{Multiturn injection for hadrons}
In the case of multiturn injection for hadrons, phase-space painting is used to increase the total circulating intensity. The variant of H$^{-}$-injection allows injection into the same phase-space area after stripping the H$^{-}$ ions to protons. In this way the intensity is increased while keeping the emittance small.
\subsubsection{Lepton injection}
In the case of high-energy leptons, as were used for the LEP accelerator, injection precision and matching is of less concern for emittance blow-up, owing to damping with synchrotron radiation. This fact can be used to inject beams off-momentum or  at an angle and increase the intensity in the same phase-space volume after damping. These techniques are called betatron and synchrotron accumulation, respectively.

\subsection{Single-turn injection and injection equipment}\label{sec:Single-turn injection and injection equipment}
\begin{figure}
\centering\includegraphics[width=.7\linewidth]{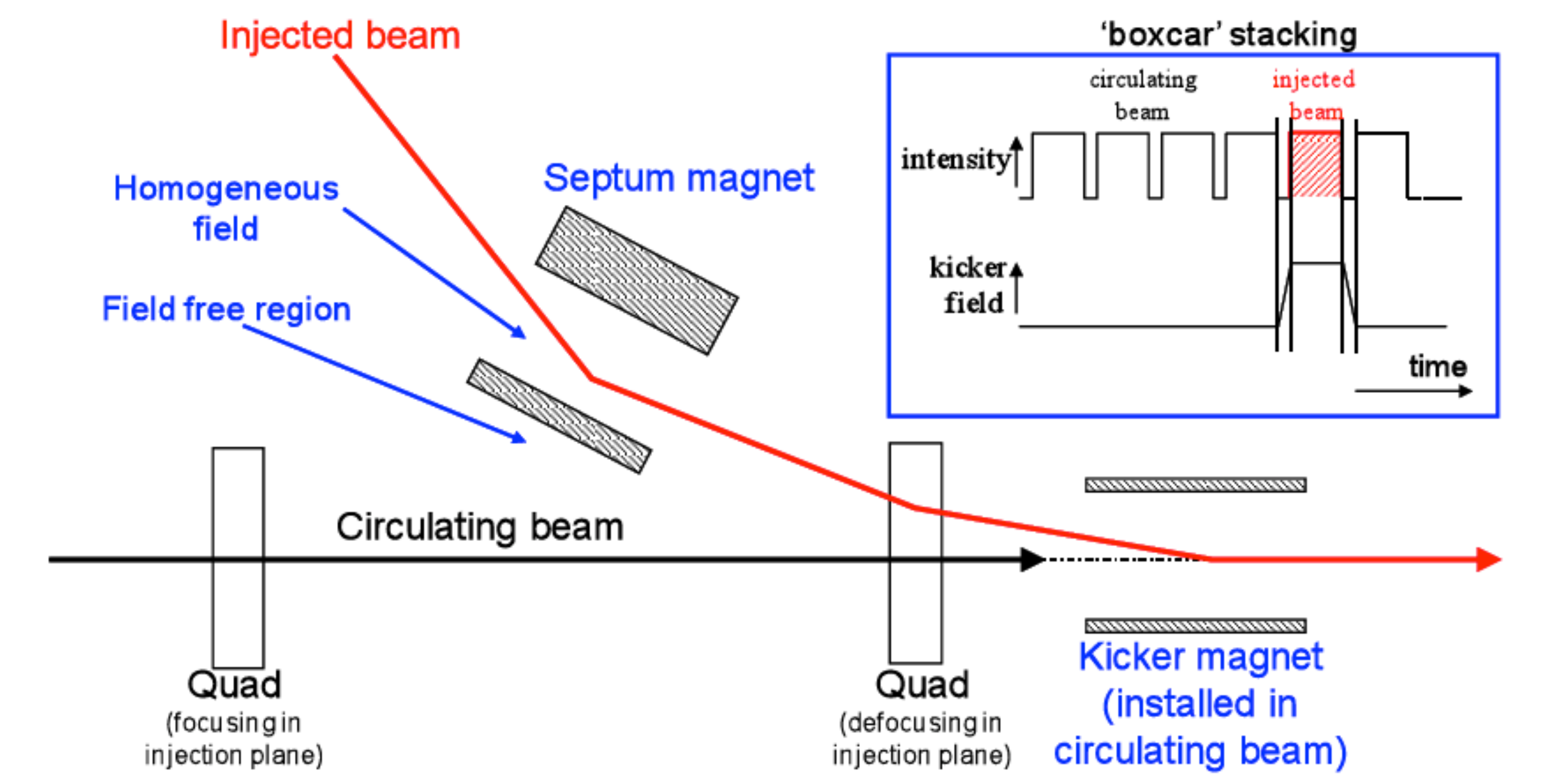}
\caption{Single-turn injection}
\label{fig:singleTurn}
\end{figure}

\begin{figure}
\centering\includegraphics[width=.9\linewidth]{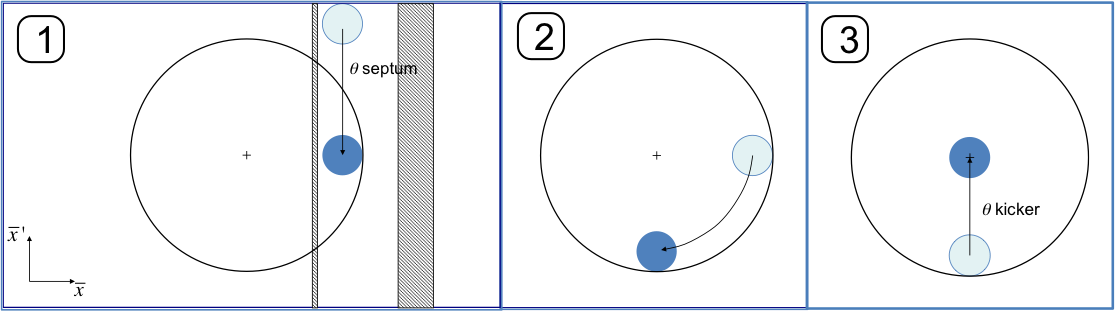}
\caption{The single-turn injection process in normalized phase-space. The beam indicated by the blue circle undergoes a large deflection as it passes through the septum magnet (plot 1), then travels through a $\pi/2$ phase advance to meet the kicker magnet (plot 2). The kicker deflection aligns the beam on the central orbit (plot 3).}
\label{fig:singleTurn_illustration}
\end{figure}

Figure  \ref{fig:singleTurn} shows the principle of single-turn injection. A transfer line brings the beam close to the orbit of the circular machine. The last two magnetic elements are special injection magnets. The first of these is called a septum magnet. It is already so close to the circulating beam that the iron yoke and coils of a conventional magnet would no longer fit between the trajectory of the injected beam and the orbit of the circulating beam. Instead, a septum magnet has an aperture with a magnetic field for the injected beam and another aperture without a magnetic field through which the circulating beam passes. Between the two regions, there is a thin septum. The septum deflects the beam onto the closed orbit at the centre of a kicker magnet. The kicker magnet compensates for the remaining angle. The kicker magnet, however, has only one aperture, through which both the injected and circulating beams pass. Kicker magnets are pulsed magnets with very fast rise and fall times, such that they provide the required magnetic field at the moment of the passage of the injected beam and no field during the passage of the circulating beam. The septum and kicker magnets are typically installed on either side of a defocusing quadrupole in the plane of the kicker magnet to minimize the kick strength. The actions of the two magnet systems in normalized phase-space are shown in Fig.\ \ref{fig:singleTurn_illustration}.

\subsubsection{Septum magnet}
A general introduction to septa can be found in Ref. \cite{bib:cas_septa}. Two main types of septa exist -- electrostatic septa and magnetic septa. Electrostatic septa have a very thin (typically ${\leq}100\Uum$) separation between the zero-field and high-field regions. They are often used for extraction systems, to minimize losses at extraction and the required kick strength or strength of the following septa.

\begin{figure}
\centering\includegraphics[width=.9\linewidth]{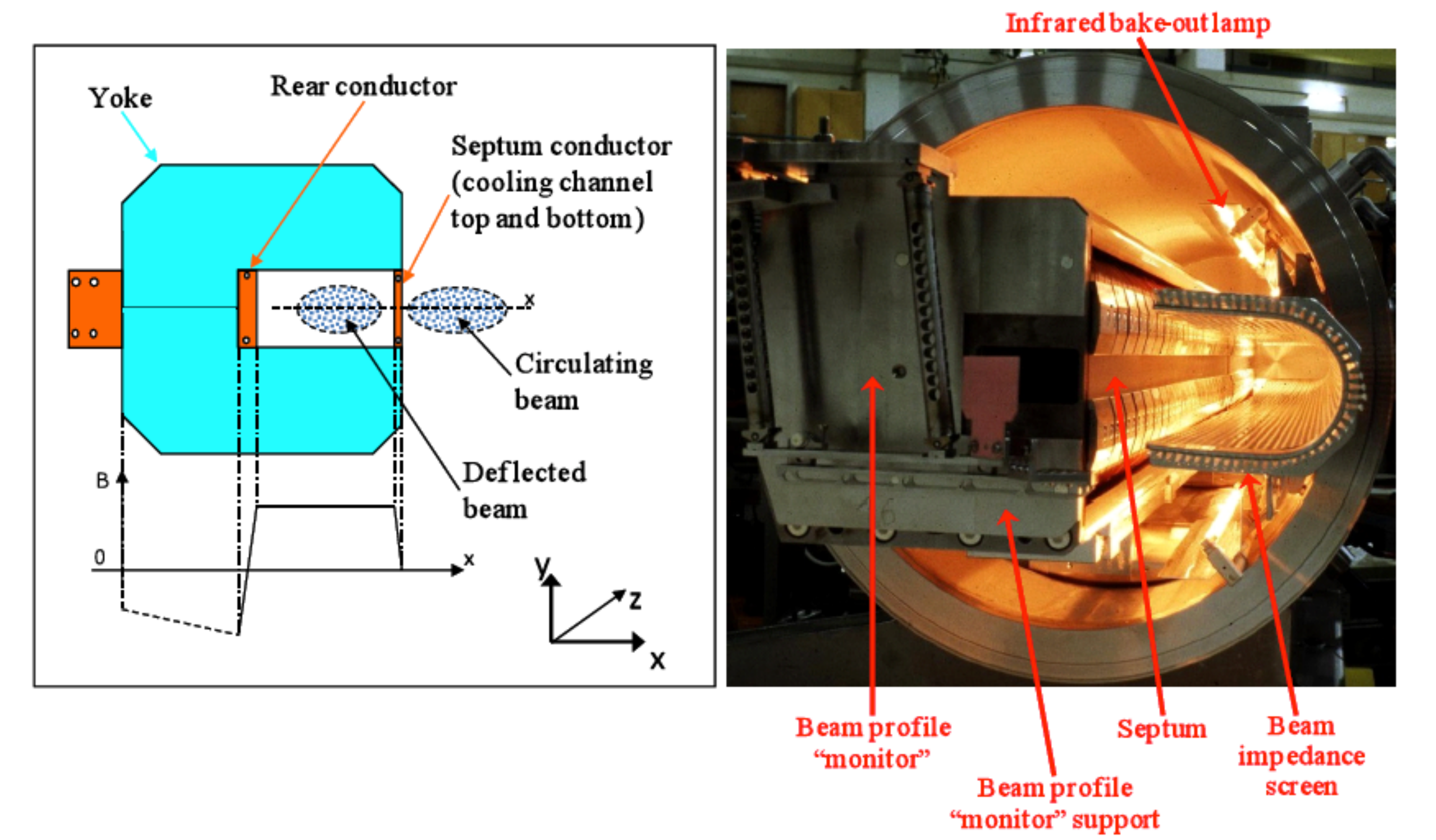}
\caption{Direct-drive pulsed septum}
\label{fig:septum}
\end{figure}

Magnetic septa are pulsed or DC dipole magnets with a 2--20\Umm\ separation between the zero-field and high-field regions.  Figure \ref{fig:septum} shows a direct-drive pulsed septum magnet. This type of magnet is often used under vacuum to minimize the distance between the circulating and deflected beams. To minimize the self-inductance and for mechanical reasons, the coil generally consists of a single turn, resulting in large required currents of typically 5--\Unit{25}{kA}. Single-turn injection uses magnetic septa.

\subsubsection{Kicker magnet}
\begin{figure}
\centering\includegraphics[width=.9\linewidth]{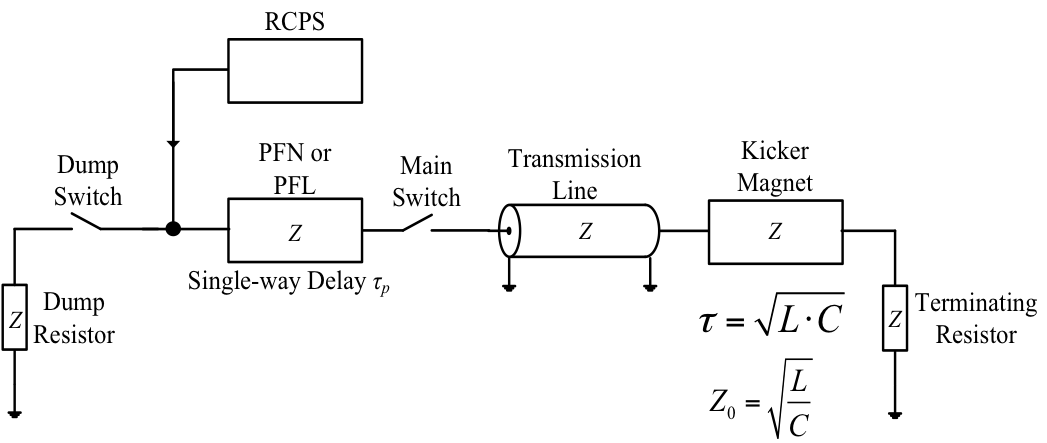}
\caption{The main sub-systems of a kicker system: The pulse forming network (PFN) or pulse forming line (PFL)  is charged to a voltage $V_p$ by the resonant charging power supply  (RCPS). When the \textit{Main Switch} closes a pulse of magnitude $V_p/2$ is launched towards the magnet through the transmission line (coaxial cable). The impedances in the system should be matched, to avoid reflections. The length of the pulse in the magnet can be controlled between 0 and 2 $\tau = \sqrt{L\cdot C}$ by adjusting the timing of the dump switch relative to the main switch. }
\label{fig:kicker_subSystems}
\end{figure}

\begin{figure}
\centering\includegraphics[width=.7\linewidth]{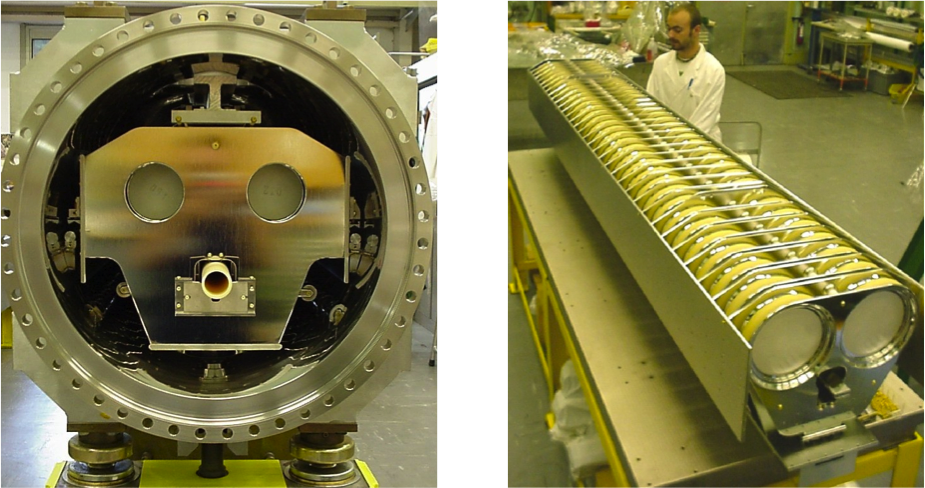}
\caption{The LHC injection kicker MKI}
\label{fig:mki}
\end{figure}

A full lecture on kicker magnets can be found in Ref. \cite{bib:cas_kickers}. Kicker magnets typically produce rectangular field pulses with fast rise and fall times. The field strength is lower than can be achieved with septum magnets. For comparison, the LHC injection septum provides roughly 12\Umrad\ of bending angle in five magnets, whereas the injection kicker magnets provide 0.8\Umrad\ in four magnets. The injection energy in the LHC is 450\UGeV. To achieve fast rise and fall times, a voltage pulse is pre-charged in a so-called pulse forming line, a coaxial cable, or a pulse forming network, all of which are lumped elements. This voltage pulse is launched towards the kicker when the \textit{main switch} is closed. The main sub-systems of a typical kicker system are shown in Fig.\ \ref{fig:kicker_subSystems}. Pulse forming networks and lines accumulate electric energy over a comparatively long time from a power supply (a few milliseconds in the case of a fast resonant charging power supply) and then release it in the form of a square pulse of relatively short duration (a few microseconds). Figure \ref{fig:mki} shows the LHC injection kicker magnet.

\begin{figure}
\centering\includegraphics[width=.5\linewidth]{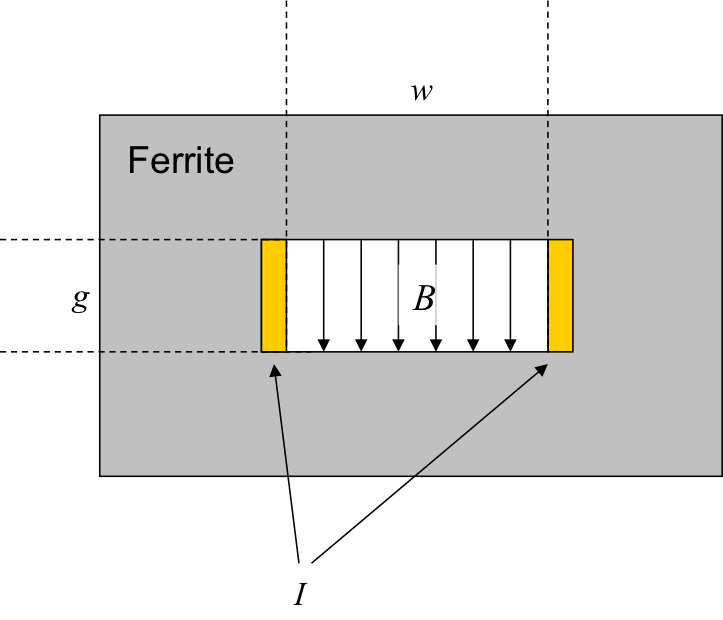}
\caption{Simplified cross-section of kicker magnet: $B$, magnetic field; $g$, gap height; $I$, current; $w$, gap width}
\label{fig:kickerCrossSection}
\end{figure}

A typical kicker cross-section is shown in Fig.\ \ref{fig:kickerCrossSection}. The characteristics  of the field in the kicker gap are:
\begin{align}
B & = \frac{\mu_0 I}{g},\\
L & = \frac{\mu_0 w l}{g},\\
\frac{\mathrm{d}I}{\mathrm{d}t}&= \frac{V}{L},
\end{align}
where $B$ is the magnetic field, $I$ the current, $g$ the gap height, $l$ the length, $w$ the gap width, $L$ the inductance, and $V$ the kicker voltage. A typical kicker $\mathrm{d}I/\mathrm{d}t$ would be \Unit{3}{kA} in 1\Uus. In general, kickers have a low inductance. Nevertheless, for short rise times, very high voltages of the order of several tens of kilovolts are required. Kicker switches, as indicated in Fig.\ \ref{fig:kicker_subSystems}, therefore need to hold high voltages and to switch high currents within short rise times. \textit{Thyratron} switches (gas tubes) or solid-state switches are frequently used for this purpose. As can be anticipated, kicker switches are associated with a number of possible failure scenarios. There are two main categories.
\begin{itemize}
\item[-] \textbf{Erratic turn-on:} The switches turn on spontaneously and the kickers fire asynchronously, \eg during the circulating beam passage instead of the injected beam.
\item[-] \textbf{Missing:} The switch does not fire.
\end{itemize}
So-called magnet \textit{flash-overs} are another kicker failure category, and are independent of the switches. A flash-over is a discharge between kicker electrodes while the kicker has high field. Depending on which cell, and hence where in the kicker, the flash-over occurs, it results in a reduced or even increased kick strength, owing to reflection.

As the bending angle exerted by the kicker magnets is generally large, kicker failures belong to the group of very fast beam loss mechanisms. If the kicker magnets are being used to inject a beam, the resulting \textit{injection oscillations} would be too large to establish a circulating beam. The beam would be lost on the first turn. In case of an asynchronous kick on the circulating beam, the resulting oscillations around the ring would be so large that the beam would be lost within a single turn. Not all kicker failures can be prevented by design but the risk for some can be minimized. For example, fast resonant charging power supplies are used to charge the pulse forming networks quickly. The charging then only takes place a few milliseconds before the kick is required. The probability of an error is reduced in this way. Many kicker failures require \textit{passive protection}, see Section \ref{sec:passive_protection}.

\subsection{Mismatch at injection}
Errors during injection can lead to beam loss due to \textit{mismatch}.
\subsubsection{Steering errors}
\begin{figure}
\centering\includegraphics[width=.9\linewidth]{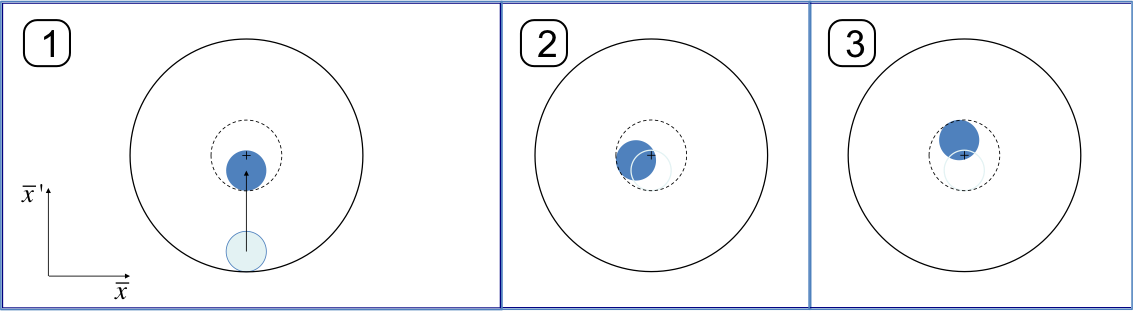}
\caption{Normalized phase-space at the injection point in the horizontal plane over three turns. The blue beam is injected with an angle error, leading to an oscillation of the particle distribution around the central orbit turn after turn.}
\label{fig:injOsc_explain}
\end{figure}

\begin{figure}
\centering\includegraphics[width=.9\linewidth]{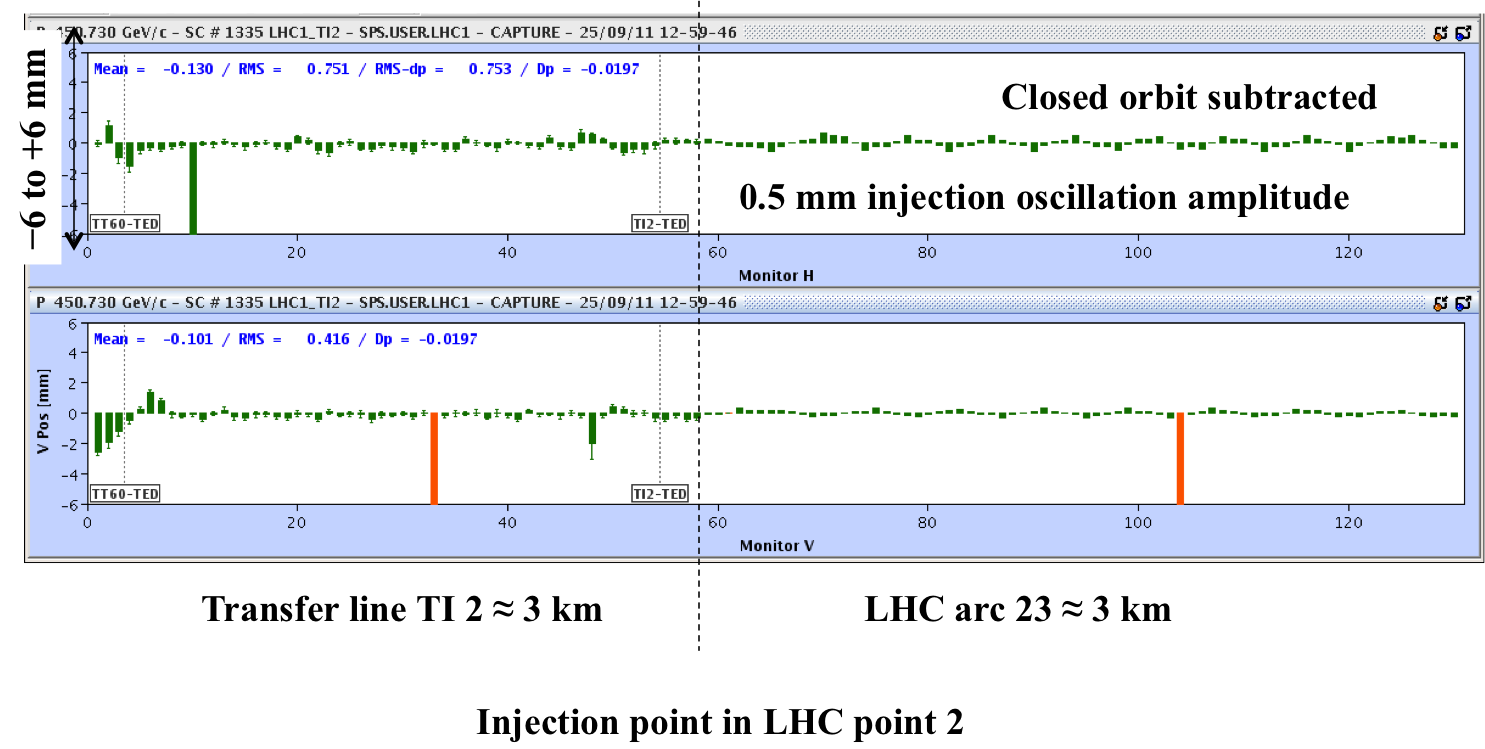}
\caption{Injection oscillation display from the LHC control room, showing the trajectory with respect to the reference in the transfer line and the first turn minus the closed orbit in the first 3\Ukm\ of the LHC.}
\label{fig:injOsc}
\end{figure}

Steering errors, a difference in position and angle (\eg $x$ and $x'$) between the circulating orbit and the injected trajectory at the injection point, lead to \textit{injection oscillations}. The injected beam will oscillate around the closed orbit. With the presence of non-linearities in the machine, this oscillation will decohere after a number of turns. As a result, the beam emittance and hence the beam sizes around the ring will increase. If the initial oscillation amplitude is large enough, it can lead to beam loss as early as the first turn. The mechanism of generating injection oscillations is illustrated in Fig.\ \ref{fig:injOsc_explain}. Figure \ref{fig:injOsc} shows the LHC injection oscillation display in the LHC control room. Instead of showing the trajectory turn after turn at one beam position monitor, the injection oscillation display shows the trajectory along many beam position monitors, covering many oscillation phases.

\subsubsection{Optical mismatch}
\begin{figure}
\centering\includegraphics[width=.5\linewidth]{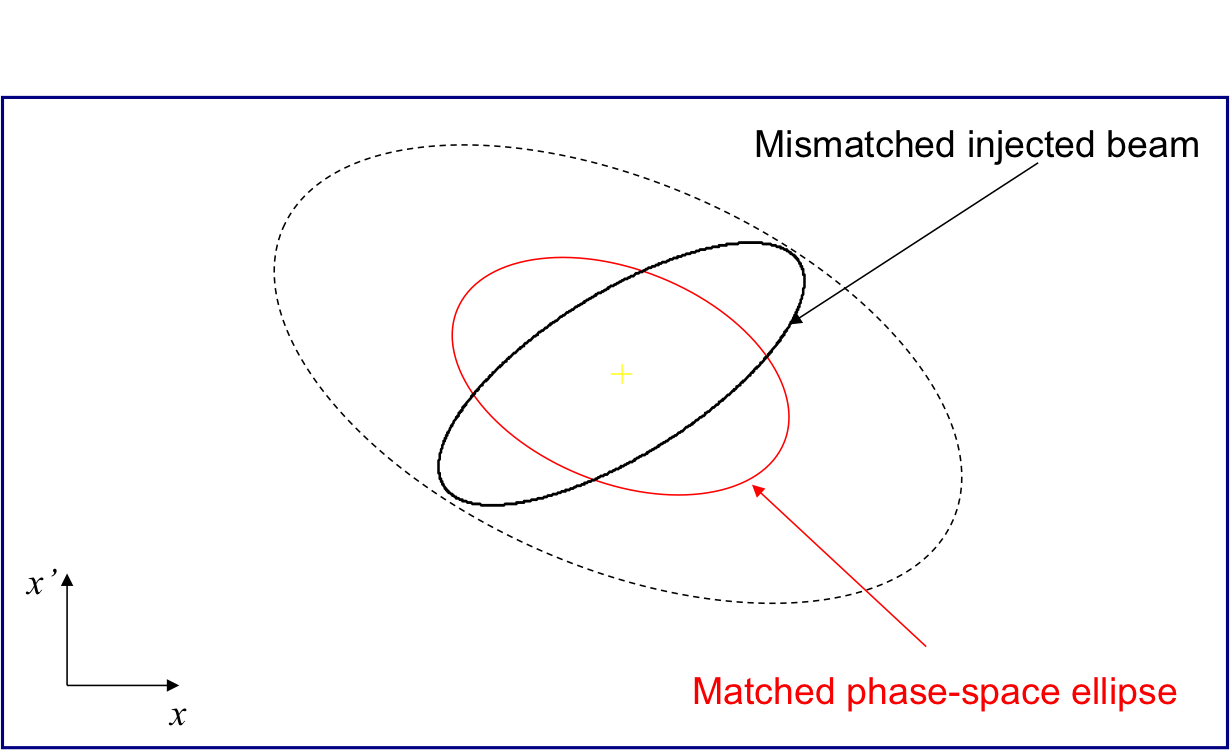}
\caption{Phase-space plot showing an optical mismatch at the injection point}
\label{fig:optMismatch}
\end{figure}

\begin{figure}
\centering\includegraphics[width=.9\linewidth]{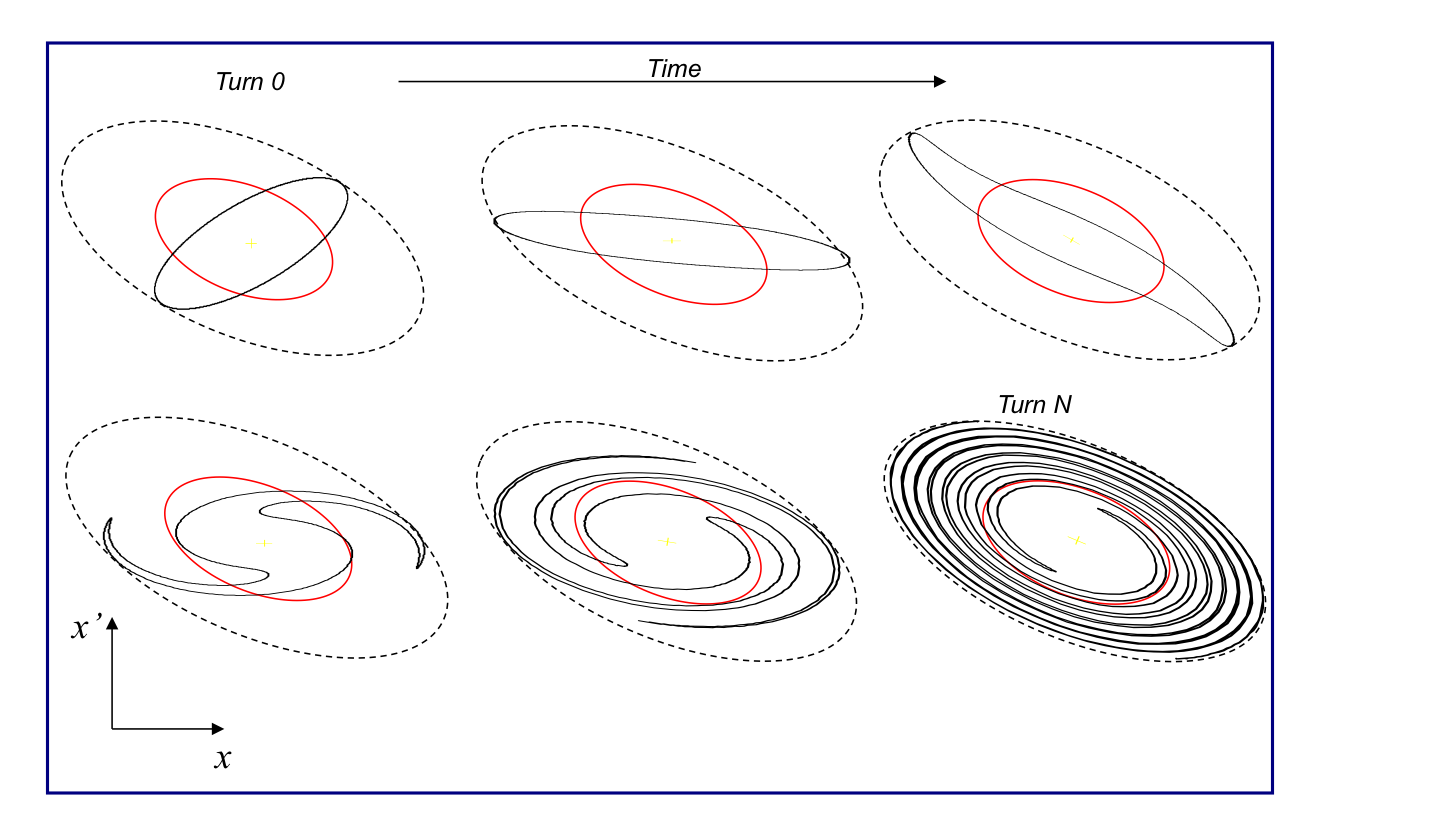}
\caption{Owing to non-linearities, optical mismatch will lead to emittance blow-up over many turns}
\label{fig:filamentation}
\end{figure}

The Twiss parameters at the injection point define the shape and orientation of the particles' phase-space ellipse. One speaks of optical mismatch if the shape and orientation of the ellipse of the injected beam does not match the ellipse of the circulating beam at the injection point.  An example of such a situation is shown in Fig.\ \ref{fig:optMismatch}. The black mismatched ellipse will change its orientation turn after turn, with its extremities following the contour of the dashed ellipse. This causes beam size beating at the injection point or any other point of observation. Non-linear effects, \eg magnetic field multipoles,  however, will introduce amplitude dependent differences in particle motion and, over many turns,  a phase-space oscillation is transformed into an emittance increase. This is illustrated in Fig.\ \ref{fig:filamentation}. So-called filamentation eventually fills a larger ellipse. The larger ellipse has the same shape and orientation as the matched one.

\section{Transfer lines}\label{sec:Transfer lines}
\begin{figure}
\centering\includegraphics[width=.9\linewidth]{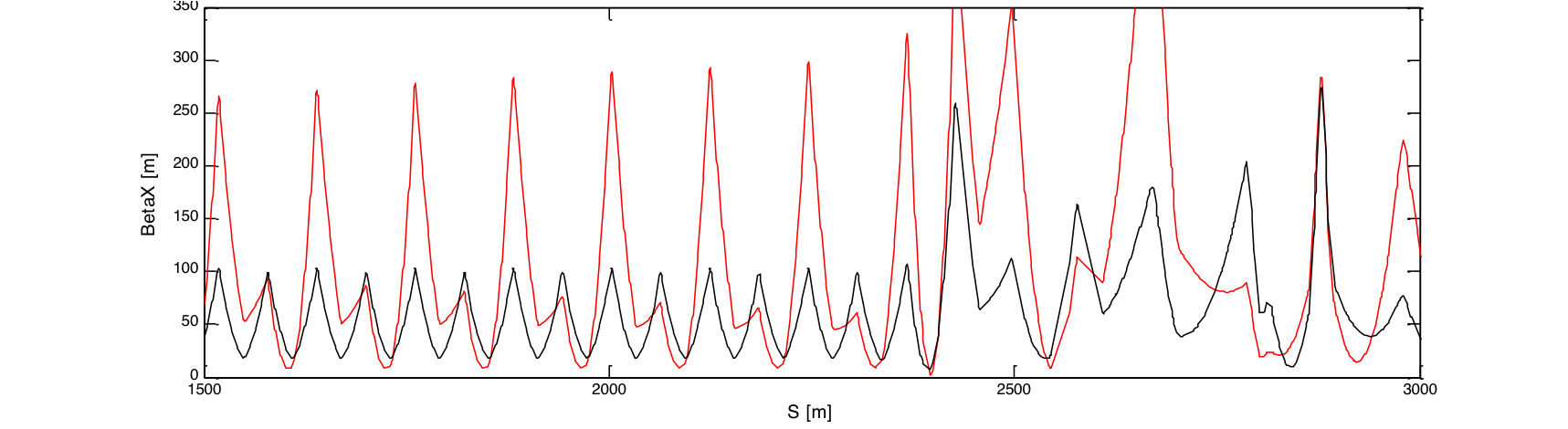}
\caption{The horizontal beta function in a transfer line as a function of the longitudinal position $s$: The strengths of the magnetic elements in the line are kept the same, but the initial conditions are modified to produce the red and black optics.}
\label{fig:tl_optics}
\end{figure}

\begin{figure}
\centering\includegraphics[width=.9\linewidth]{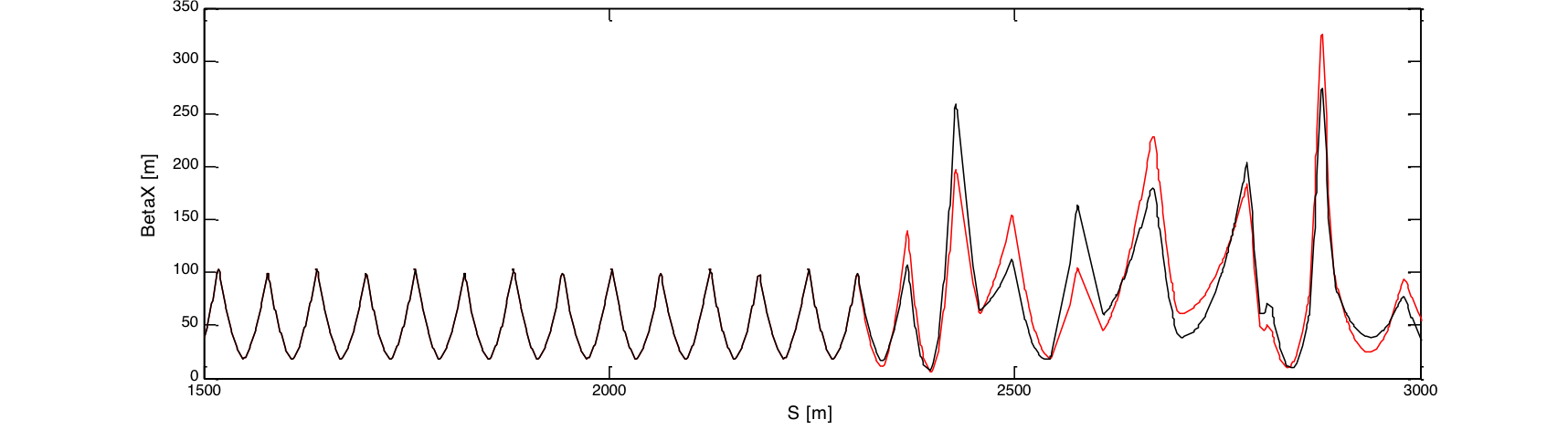}
\caption{The effect of an error in a transfer line is only seen downstream of the error location. For the red beta function, a quadrupole error was introduced at location $s \approx 2300\Um$.}
\label{fig:tl_error}
\end{figure}

Transfer lines transport the beam between accelerators and onto targets, dumps, and instruments. The beam dynamics in a transfer line are not only determined by the strength of the elements in the transfer line. Instead, the Twiss parameters at any point in the line, $\alpha$(s), $\beta$(s), are also a function of the initial functions, $\alpha_1$ and $\beta_1$. This is illustrated in Fig.\ \ref{fig:tl_optics}. The strengths of the line are kept the same but the initial conditions are modified. As a result, the beta functions change. The same is true for the dispersion function.
Another difference with respect to circular machines is the effect of an error at a particular location in the line. Whereas in a circular machine the error will introduce a perturbation around the entire ring, in a transfer line the error will affect only the part of the line downstream of the error location, see Fig. \ref{fig:tl_error}.

If transfer lines are used to link two machines, the Twiss parameters of the extraction point have to be propagated through the line and matched to the Twiss parameters at the injection point of the next machine by adjusting the quadrupole strengths in the line accordingly, see Fig. \ref{fig:tl_matching}. From the previous discussion, we know that the Twiss parameters, and hence the orientation of the phase-space ellipse of the injected beam, have to be the same as the Twiss parameters,  and orientation, of the phase-space ellipse of the circulating beam at the injection point, to avoid optical mismatch.  The Twiss parameters can be propagated if the transfer matrix $M_{1\to 2}$ is known:
\begin{figure}
\centering\includegraphics[width=.7\linewidth]{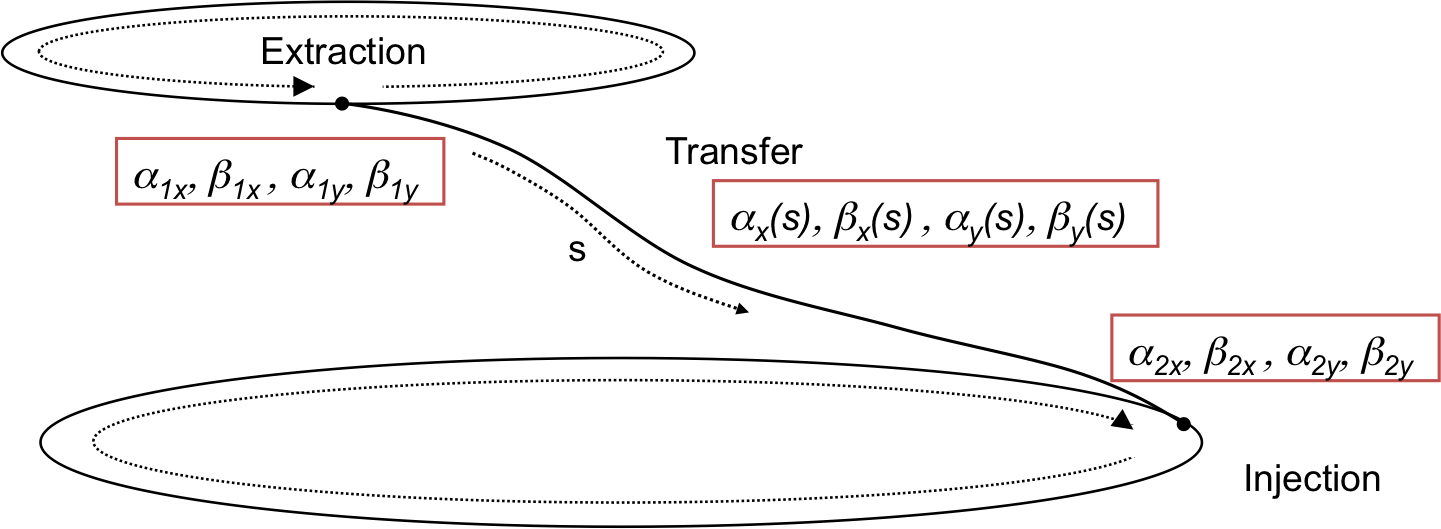}
\caption{If transfer lines are used to link two machines, the Twiss parameters of the extraction point have to be propagated through the line and matched to the Twiss parameters at the injection point of the next machine.}

\label{fig:tl_matching}
\end{figure}
\begin{equation}
\begin{pmatrix}
x \\ x'
\end{pmatrix}_{s_2}
= M_{1 \to 2}
\begin{pmatrix}
x\\ x'
\end{pmatrix}_{s_1} =
\begin{pmatrix}
C & S\\
C' & S'
\end{pmatrix}\cdot
\begin{pmatrix}
x\\ x'
\end{pmatrix}_{s_1}\ ,
\end{equation}
\begin{equation}
\begin{pmatrix}
\beta_2\\
\alpha_2\\
\gamma_2
\end{pmatrix}
=
\begin{pmatrix}
C^2 &  -2CS & S^2\\
- CC'& CS'+SC'& -SS'\\
C'^2 & -2 C'S' & S'^2
\end{pmatrix}
\cdot
\begin{pmatrix}
\beta_1\\
\alpha_1\\
\gamma_1
\end{pmatrix}
\ .
\end{equation}
Typically, eight variables need to be matched ($\alpha$, $\beta$, the dispersion $D$ and its derivative $D^\prime$ in both planes). Frequently, constraints such as phase advance requirements for transfer line collimators or insertions for special equipment such as stripping foils have to be respected. Independently powered quadrupole magnets with adjustable strength are used for this purpose. If the transfer lines are long, the problem can be simplified by designing the lines in separate sections. The main part of the line then consists of a regular central section with a regular lattice, \eg a FODO or doublet with the quadrupoles at a regular spacing and powered in series. At the end and the beginning of the line, initial and final matching sections with independently powered quadrupoles are installed. An example of this type of layout is the LHC transfer line shown in Fig.\ \ref{fig:lhc_tl}.\\
\begin{figure}
\centering\includegraphics[width=.79\linewidth]{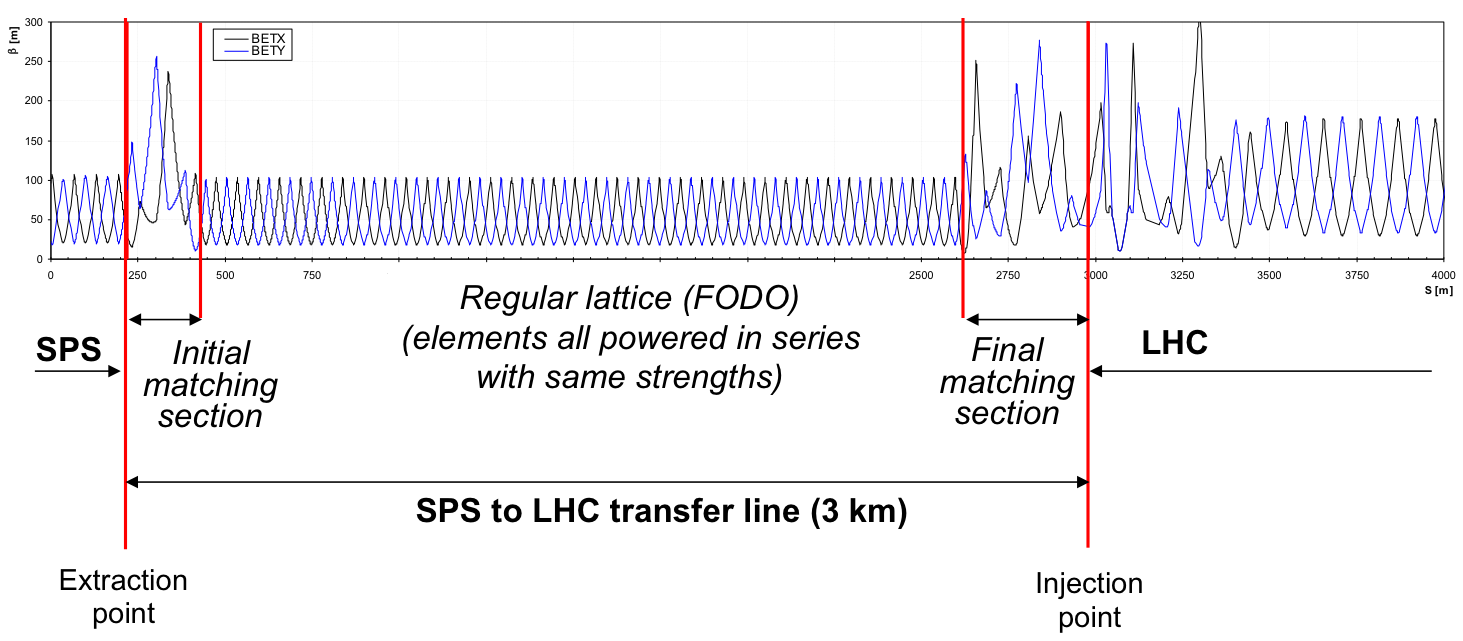}
\caption{Layout of the SPS to LHC transfer line with a long central section with a regular lattice and initial and final matching sections.}
\label{fig:lhc_tl}
\end{figure}

To counteract trajectory deviations due to magnet misalignments and field and powering errors, transfer lines are usually also equipped with beam position monitors and independently powered small dipole corrector magnets. Horizontal correctors and beam position monitors are located at large $\beta_x$ and vertical correctors and beam position monitors are located at large $\beta_y$.

\section{Extraction}\label{sec:Extraction}
As in the case of injection, there are various different techniques to extract the beam from a circular machine. A few methods will be briefly summarized.
\subsection{Single-turn fast extraction}
\begin{figure}
\centering\includegraphics[width=.7\linewidth]{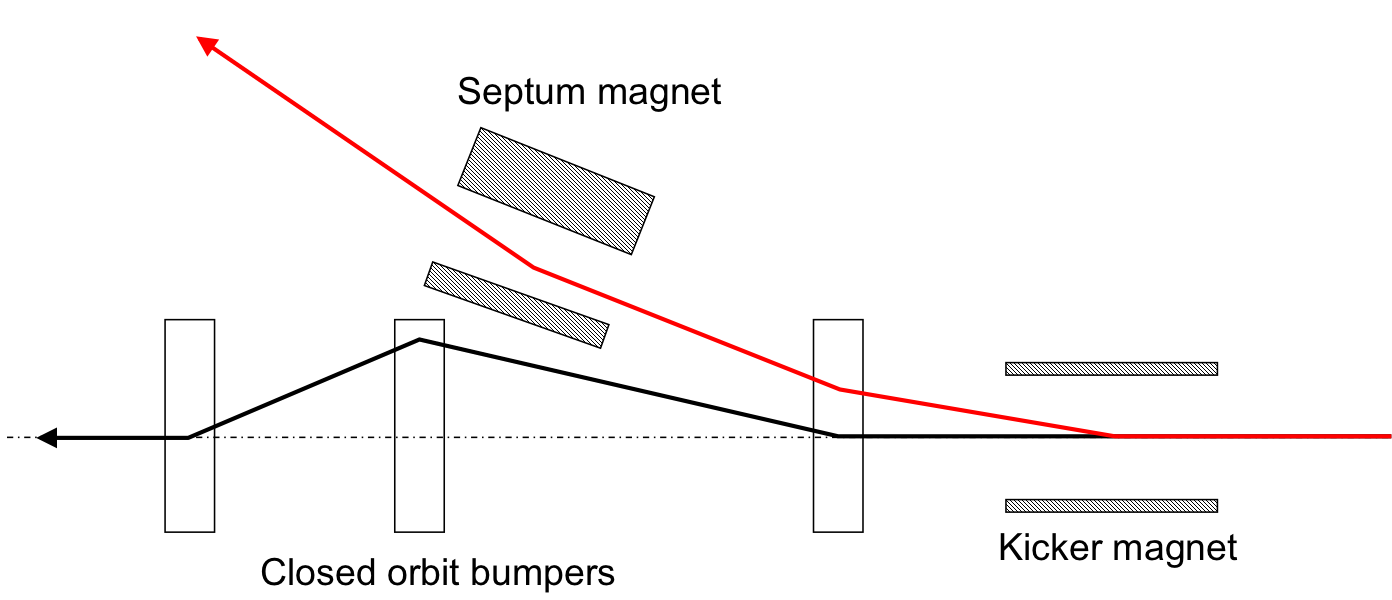}
\caption{Travelling from right to left, the beam is deflected by the kicker magnet into the septum, which bends the beam into the transfer line.}
\label{fig:single_turn_extraction}
\end{figure}

\begin{figure}
\centering\includegraphics[width=.7\linewidth]{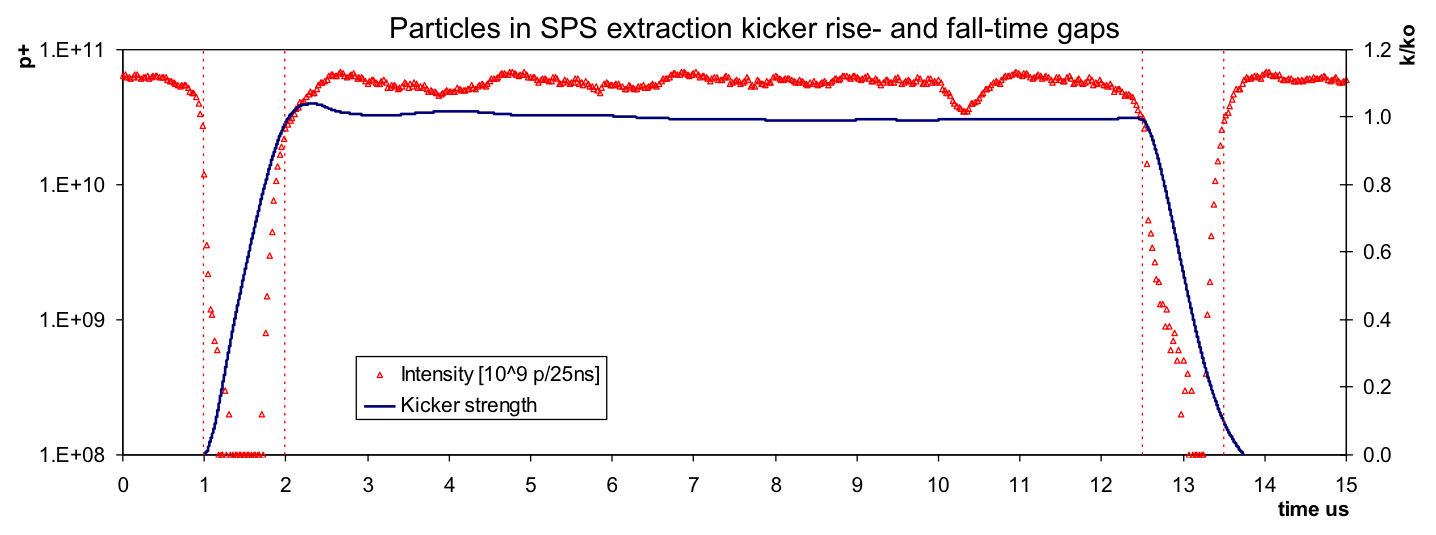}
\caption{One turn takes 23\Uus\ for the SPS beam at 400\UGeVc. For the CERN Neutrino to Gran Sasso beam, two batches, each of 10\Uus, fill the SPS. The batches are extracted one after the other at intervals of 50\Ums. The extraction kickers have to rise between the batches. The beam in the gaps between the batches is lost  on the extraction septum blade or in the SPS.}
\label{fig:cngs_extraction}
\end{figure}

As for single-turn injection, kicker magnets and septa are used for single-turn fast extraction. To reduce the required kicker magnet strength, the beam is sometimes moved close to the septum  by a closed orbit bump. An example of a typical fast extraction layout is shown in Fig.\ \ref{fig:single_turn_extraction}. The kicker magnet deflects the entire beam into the septum in a single turn. The septum then bends the beam into the transfer line. The smallest  deflection angles are required for  a phase advance of $\pi/2$ between the kicker and the septum. The septum deflection may also be in the other plane from the kicker deflection. Losses around the ring or in the extraction channel at the moment of the extraction process can originate from orbit errors (\eg closed orbit bump errors), kicker failures, kick synchronization errors or particles  in the extraction gap during which the extraction kicker field rises (\eg uncaptured beam).  An example of a kicker rising in an extraction gap populated with particles is shown in Fig.\ \ref{fig:cngs_extraction} for the CERN Neutrino to Gran Sasso fast extraction from the CERN SPS.

\subsection{Non-resonant multiturn extraction}
This type of extraction is used to fill a larger machine from a smaller one over several turns. The beam is sliced into equal parts by a fast orbit bump that puts the whole beam onto the septum. The beam is extracted in a few turns with the machine tune rotating the beam. This type of extraction is an intrinsically high-loss process.

\subsection{Resonant multiturn extraction}
Multipole fields from sextupoles or octupoles are used to distort the circular normalized phase-space and define a stable area delimited by unstable fixed points. The beam is slowly extracted by approaching the respective resonance increasingly closely  and going across it. By adjusting the tune change speed, the extracted intensity can be controlled and a constant particle spill can be achieved.
\subsection{Resonant low-loss multiturn extraction}
An alternative to the non-resonant multiturn extraction with lower losses is the resonant low-loss multi\-turn extraction. Non-linear fields (sextupole and octupole) are used to create islands of stability in phase-space. The particles are driven into the islands through tune variations. The islands are separated further in phase-space by  varying the non-linear fields. The islands are then extracted in a similar manner as for the non-resonant multiturn extraction with a fast bump turn after turn.

\section{Machine protection for beam transfer}
The typical concepts for machine protection systems for beam transfer will be introduced with the example of the SPS to LHC injection process and the LHC beam dumping system. The LHC is filled with 12 injections from the SPS. The extraction system in the SPS is a fast extraction using a large closed orbit bump of ${\approx}35\Umm$ amplitude in the horizontal plane and horizontal kickers and septa. The LHC transfer lines are $3\Ukm$ long. The LHC injection system is a single-turn injection consisting of  horizontal septa magnets and vertical kicker magnets that deflect the beam on the LHC closed orbit. The LHC beam dump system is a fast extraction system consisting of a horizontal kicker system and vertical septum magnets. The beam is deflected into a $900\Um$ dump channel, at the end of which the beam dump block is located.

The design parameters of the LHC proton beams in the SPS at the moment of extraction and after acceleration in the LHC are summarized in Table \ref{tab:lhc_param}. The equipment damage limit has been calculated and measured to be of the order of $100\UJ/\UcmZ^3$, which corresponds to only $\approx5\%$ of the injected intensity in the LHC.

\begin{table}
\caption{LHC nominal beam parameters}
\label{tab:lhc_param}
\centering
\begin{tabular}{lrrcd}  
  \hline\hline
   &  \multicolumn{1}{r}{\textbf{Energy}}  &  \multicolumn{1}{r}{\textbf{Number of bunches}}  &  \multicolumn{1}{c}{\textbf{Bunch intensity}}  &  \multicolumn{1}{c}{\textbf{Stored energy}}  \\
  &  \multicolumn{1}{r}{\textbf{[\UGeVZ]}}  &     &     &  \multicolumn{1}{c}{\textbf{[\UMJZ]}}  \\
 \hline
 \multicolumn{1}{l}{SPS extraction}  &   450   &  288   &  \multicolumn{1}{c}{$1.7 \times 10^{11}$}  &  2.4  \\
 \multicolumn{1}{l}{LHC top energy}  &   7000   &   2808   &  \multicolumn{1}{c}{$1.7 \times 10^{11} $}  &   360  \\
\hline\hline
\end{tabular}
\end{table}

The basics of injection and extraction protection can be summarized as:
\begin{itemize}
\item \textbf{Ensure correct settings:} \eg energy tracking systems, power supply surveillance;
\item \textbf{Ensure kicker synchronization:} particle-free gaps (\ie gap cleaning), locking extraction kicker triggers to RF revolution frequency, \etc
;\item \textbf{Provide passive protection system:} in the form of collimators and absorbers against fast possible failure mechanisms and unavoidable kicker failures.
\end{itemize}

\subsection{Monitoring, permits, and kickers}
Different interlocking systems are installed in the SPS, the LHC, the LHC injection region, and the SPS extraction region and transfer lines. A number of systems are monitored and a Boolean status from these systems is transferred to the interlock system. The result of the combination of the different inputs is called a \textit{permit}, \ie injection permit, extraction permit, SPS beam permit, \etc
These permits are each connected to a kicker system. The SPS beam permit and LHC beam permit are connected to the dump kicker systems. If the beam permit becomes \textit{false}, the dump kickers fire and dump the beam on the respective beam dump blocks. Without an extraction permit in the SPS, the beam is not extracted from the SPS. The kickers will not fire. Similarly, the LHC injection kickers will not fire without an injection permit.

\begin{figure}
\centering\includegraphics[width=.8\linewidth]{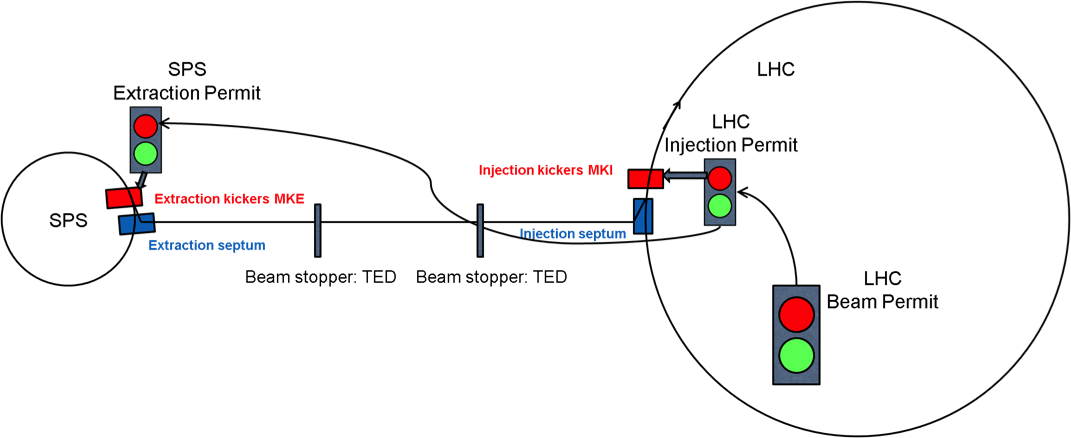}
\caption{Layout of the SPS to LHC transfer with the extraction kicker MKE in the SPS, the transfer line (TT60 and TI2) and injection kicker MKI in the LHC, with corresponding extraction and injection permits.}

\label{fig:tl_interlocking}
\end{figure}

\begin{figure}
\centering\includegraphics[width=.7\linewidth]{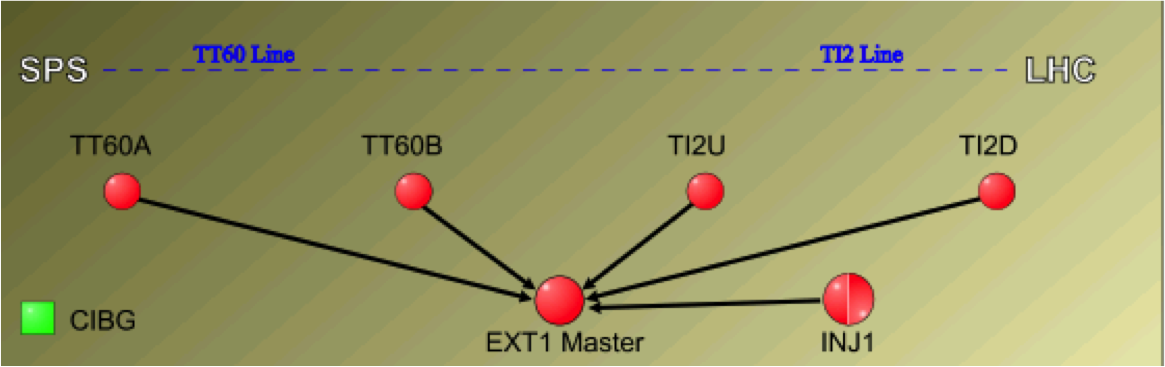}
\caption{Monitoring systems associated with equipment in the different areas involved in the SPS to LHC transfer are combined into sub-results, which are input to the extraction permit. Each circle in the drawing represents a `beam interlock controller'. INJ1 is the injection permit of LHC beam 1.}
\label{fig:permit_hierarchy}
\end{figure}

All permits involved in the SPS to LHC transfer are connected hierarchically. The beam from the SPS must not be extracted if the LHC injection permit is not \textit{true}. The SPS extraction permit therefore depends on the LHC injection permit. Also, the beam should not be injected into the LHC if the LHC is not ready (\ie if the LHC beam permit is \textit{false}). The LHC beam permit is input to the LHC injection permit. Figure \ref{fig:tl_interlocking} shows an overview of the permit hierarchy of the SPS to LHC transfer. The hardware systems in the SPS and LHC to which the different monitoring systems are connected and which provide permits are LHC-type beam interlock controllers (BICs) \cite{bib:BIC}. The interlocks of equipment of the different areas in the transfer lines are combined  in local BICs (\eg TT60A, TI2U for transfer lines TI 2 and TT60, respectively), see Fig. \ref{fig:permit_hierarchy}.

\subsubsection{Which systems are monitored?}
Almost every element in the SPS to LHC transfer regions is monitored and interlocked:
\begin{itemize}
\item the orbit at the SPS extraction point, measured using a number of dedicated beam position monitors;
\item power supply currents, measured $2\Ums$  before SPS extraction for all extraction and transfer line circuits, including trajectory correcting dipole magnets;
\item septa currents, kicker state, and kicker charging voltage;
\item vacuum valves and optical transition radiation (OTR) screens, which both have to be out for high-intensity beams; \item the settings and gaps of passive protection devices;
\item the beam loss reading of the beam loss monitors -- if one shot generates losses above threshold, the next shot will be inhibited;
\item the trajectory via the reading of beam position monitors.
\end{itemize}
Figure 20 shows the graphical user interface display for the beam interlock controllers installed in the first part of the LHC transfer line TI 8, called TT40.

\begin{figure}
\centering\includegraphics[width=.7\linewidth]{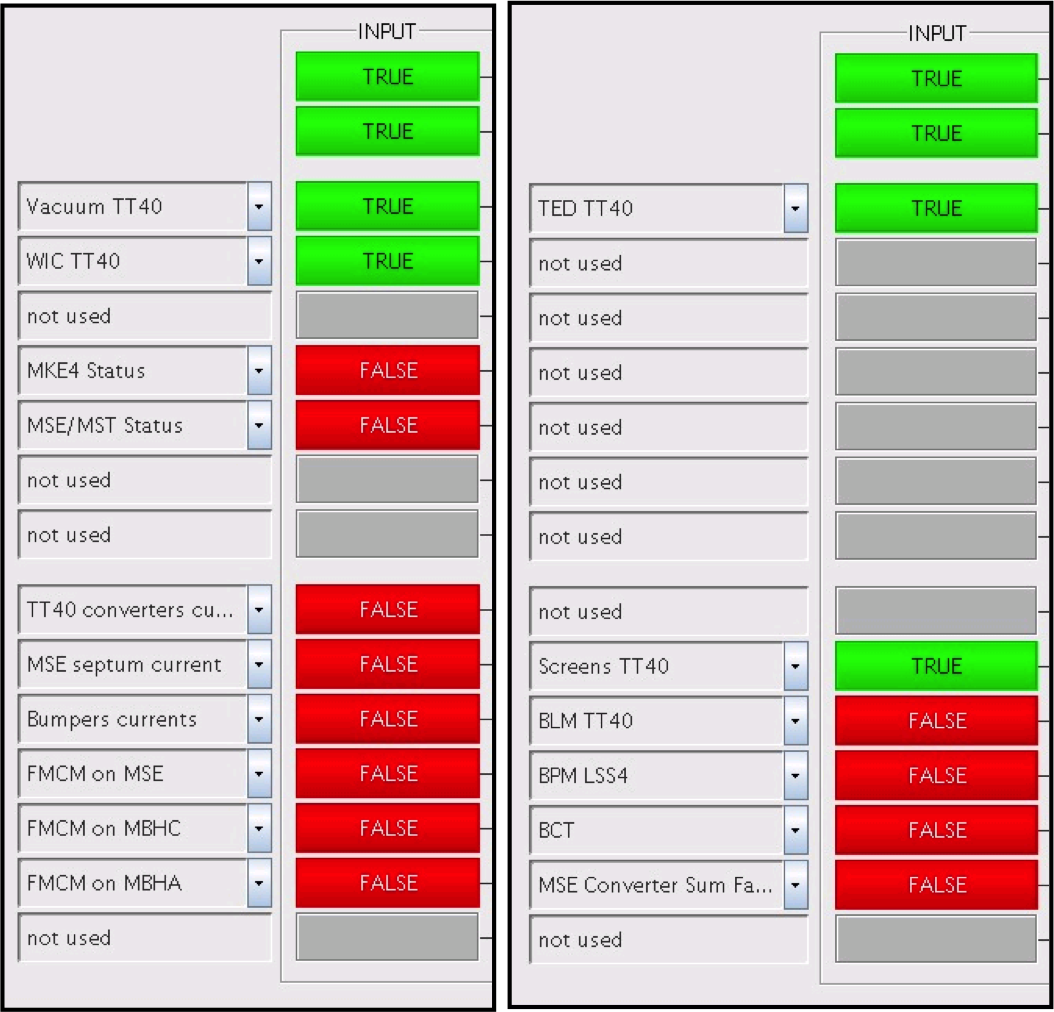}
\caption{Graphical user interface displays of two beam interlock controllers in TT40, the upstream part of transfer line TI 8. The vacuum system, the warm magnet interlock system (WIC), the extraction kicker MKE4 state, the septa MSE and MST states, the TT40 power supply currents, the septa currents, the orbit bump corrector magnet currents, the fast magnet current change monitors (FMCM), which survey fast changes in case of failures instead of absolute current values, the optical transition radiation (OTR) screen positions, the beam loss monitors (BLM) and the extraction orbit with BPM LSS4 are all input to these interlock controllers.}
\label{fig:tt40_bics}
\end{figure}
\subsubsection{Masking of interlocks}
\begin{figure}
\centering\includegraphics[width=1\linewidth]{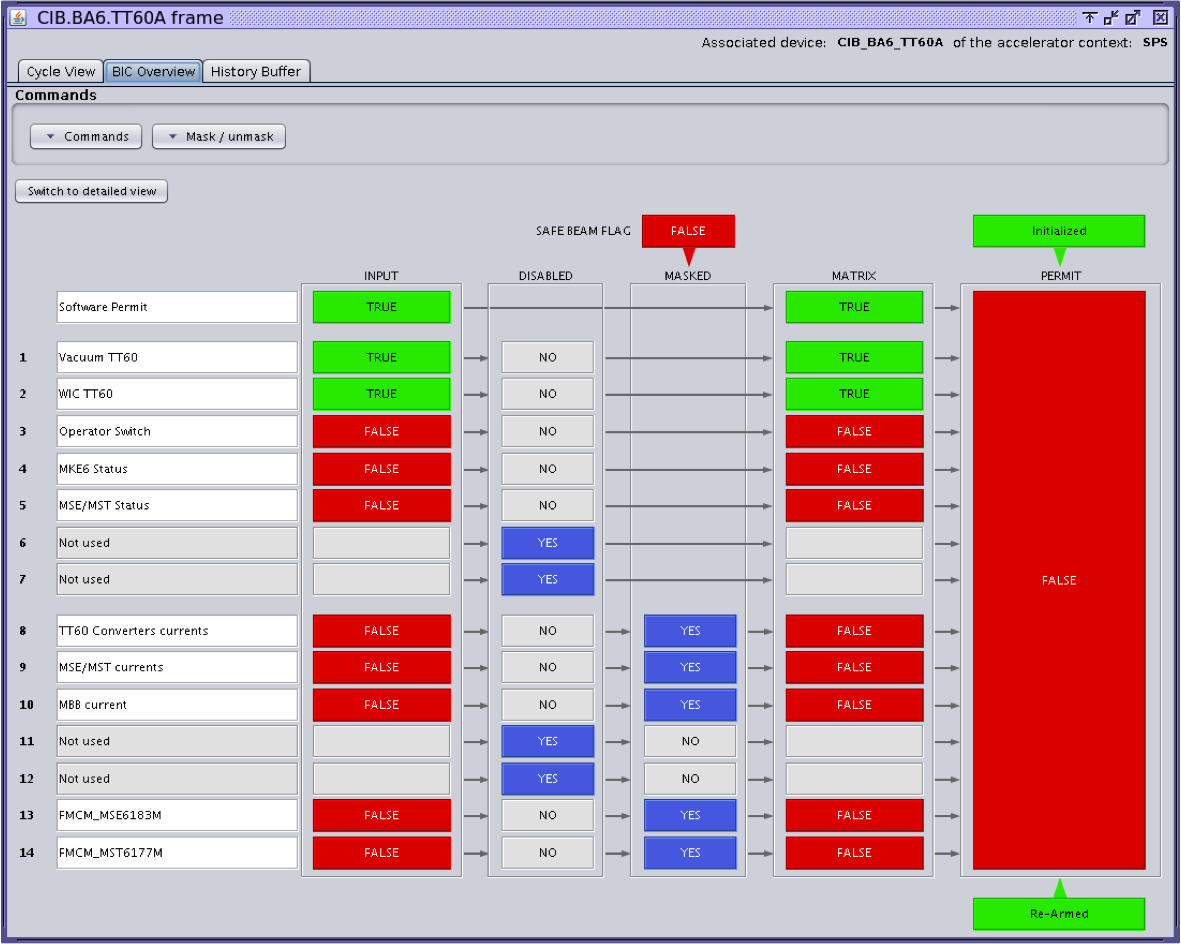}
\caption{Graphical user interface display of a beam interlock controller in TT60, which is the upstream part of LHC transfer line TI 2.  Inputs 8--14 are maskable and can be disabled in case of intensity below the \textit{set-up beam flag}. In this example, the set-up beam flag is \textit{false} and the masks do not have an effect, \ie in the column ` matrix' the masked inputs return \textit{false} (red), despite the mask.}
\label{fig:bic_masks}
\end{figure}

Appropriate settings  (\eg of corrector magnet power converters or collimators that have to be aligned around the beam trajectory) can frequently only be established with the beam. For the first set-up, these types of interlock should, therefore, not be taken into account when generating permits. In LHC jargon, the disabling of interlocks is called `masking'. A number of systems can have maskable interlocks in the beam interlock controller matrix.
For the LHC-type beam interlock controllers, masking is only allowed at low intensity, the so-called \textit{set-up intensity}. The set-up intensity is derived from the SPS or LHC beam current transformers and is distributed across the machines in the form of the \textit{set-up intensity flag}. A dedicated distribution system has been put in place --  the \textit{Safe Machine Parameter System} \cite{bib:smp}. The interlock masks are automatically ignored if the measured intensity is above the set-up intensity limit. Figure \ref{fig:bic_masks} shows the graphical user interface of one of the beam interlock controllers involved in the SPS to LHC transfer. A number of inputs are maskable. Several masks have been set. As the set-up beam intensity flag is \textit{false}, the masked interlocks are nevertheless taken into account.

\subsubsection{Set-up intensity and other concepts}
The set-up intensity has to be reasonably high, for testing under representative conditions, but must be below the equipment damage limit. For the LHC, the damage limit has been simulated and measured as ${\approx}2 \times 10^{12}$ protons at 450\UGeV\, with a normalized emittance of 3.5\Uum \cite{bib:tt40_experiment}. Based on this limit, the set-up intensity has been defined as $1 \times 10^{12}$ protons at injection energy. Its energy dependence has been obtained from simulation:
\begin{equation}
\left( \frac{E\ [\UGeV]}{450\ [\UGeV]} \right)^{1.7} \times I\ [\mathrm{p}] \leq 1\times 10^{12} .
\end{equation}

\begin{figure}
\centering\includegraphics[width=.5\linewidth]{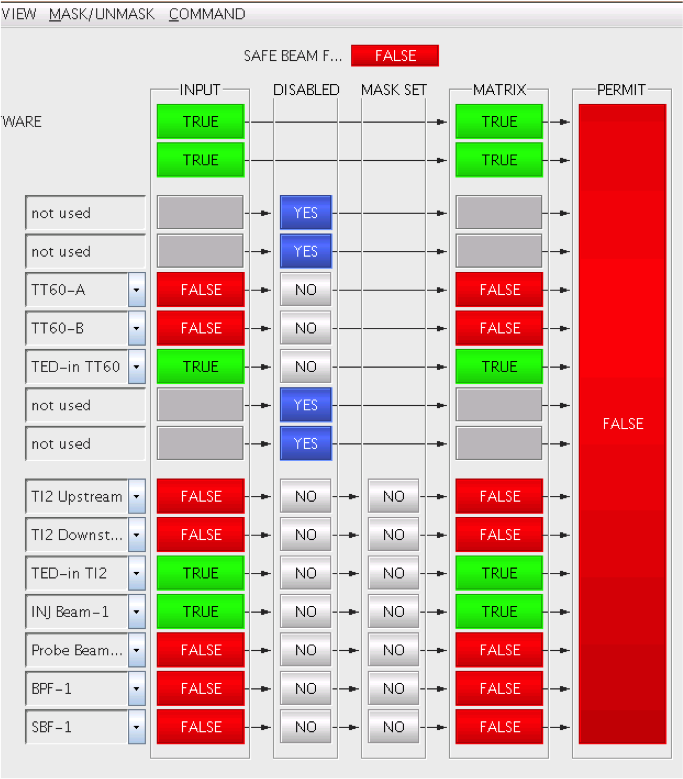}
\caption{Graphical user interface of the beam interlock controller generating the SPS extraction permit for LHC beam 1. At the very top, the SPS set-up beam flag state can be seen, labelled `Safe Beam Flag I'. The last three inputs to the controller are the \textit{SPS probe beam flag}, the \textit{LHC beam presence flag} and the \textit{LHC set-up beam flag}.}
\label{fig:bic_flags}
\end{figure}

The LHC injection protection concept cannot rely only on monitoring systems and interlocking monitored systems. The LHC consists of thousands of sub-systems, which all have to be in the correct state to receive the beam. Only a fraction of these are directly connected to the interlock system. The SPS extraction permit for a high-intensity beam is therefore only given if there is already a beam circulating in the LHC. The so-called \textit{probe intensity} has to be injected first, to check whether everything really is ready for the beam. The \textit{probe intensity} is  $ {<}5 \times 10^{9}$ protons, almost a factor of 100 below the intensity of 1 nominal bunch. The  \textit{probe intensity} information is generated from the SPS beam current transformers and the \textit{beam presence} information is generated from the sum signal of the beam position monitors in the LHC. Both are distributed in the form of flags over the \textit{Safe Machine Parameter System}.
The condition that is hard-coded in the beam interlock controller that generates the SPS extraction permit, Fig. \ref{fig:bic_flags}, is
\begin{equation}
\text{SPS probe beam flag} \lor \{\text{beam presence} \land [\lnot (\text{LHC set-up beam flag}) \lor \text{SPS set-up beam flag} ]\}\ .
\label{eq:condition}
\end{equation}
This condition translates as: `high-intensity extraction from the SPS above the \textit{probe intensity} is only allowed if there is a beam circulating in the LHC'. The set-up beam flag in the LHC has to be forced to \textit{false} to remove potential masks if the SPS intensity is above the \textit{set-up intensity}.

\subsection{Protection against fast failures}\label{sec:passive_protection}
\begin{figure}
\centering\includegraphics[width=.7\linewidth]{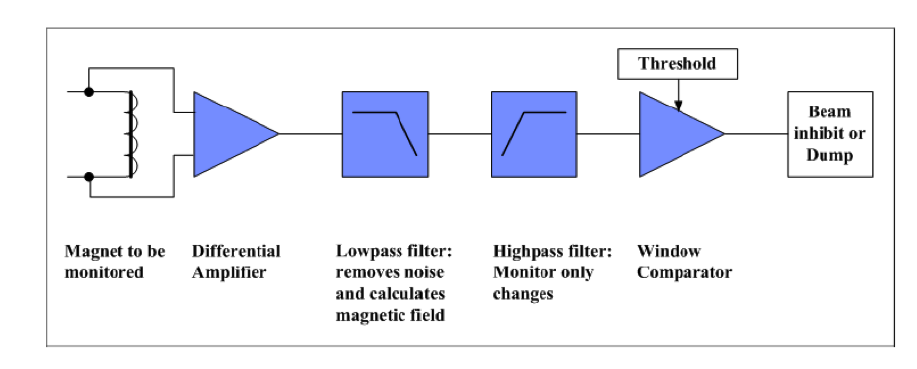}
\caption{The fast current change monitor measures the voltage of a power supply and calculates real-time current changes using a low-pass filter followed by a high-pass filter. The low-pass filter removes the noise and calculates the current; the high-pass filter removes the d.c. part and shows only current changes.}

\label{fig:fmcm}
\end{figure}

Beam transfer using fast extraction systems or single-turn injection cannot be aborted once the beam transfer is triggered. All involved systems have to be checked and give the OK before the trigger arrives. For the power supplies in the SPS to LHC transfer, the last current verification happens  ${\approx}2\Ums$ before  the extraction kickers fire. Some of the circuits, however, have very low time constants for the current decay in the case of a power supply failure. The magnetic field can, therefore, still be significantly wrong at the moment of the beam passage, even if the power supply only fails at the very last moment. An example is the extraction septum circuit, with a time constant of only $23\Ums$ and large bending angles of $12\Umrad$. The trajectory would be changed by roughly 40$\sigma$ within $1\Ums$ in the case of a power supply failure. The aperture in the transfer line is only 10$\sigma$. Instead of relying only on the classic power converter current surveillance, a faster system had to be put in place  for the septa circuits and several other circuits. This system is the so-called \textit{fast magnet current change monitor} (FMCM). Instead of measuring small currents, it measures changes in the voltage of a critical power supply. It can detect relative current changes of $10^{-4}$ in less than $1\Ums$. It was successfully deployed for septa current surveillance, with a reaction time of 50\Uus\ for a current change of $0.1\%$ \cite{bib:fmcm}. Figure \ref{fig:fmcm} shows the FMCM system.

Kicker failures can develop on microsecond time-scales. No surveillance system would be fast enough to stop the beam in time. If a failure occurs during the kick pulse, other means of protecting downstream equipment from the beam impact have to be envisaged. These means are passive protection systems, in the form of absorbers and collimators.

\subsubsection{Passive protection}
The design criteria for a passive protection system can be obtained by answering a number of questions.
\begin{itemize}
\item Against which failure should the passive protection devices protect? This will identify where to put the devices and how many will be needed. Examples will be given later on.
\item What is the damage level, quench level, or allowed radiation level of downstream equipment? This will define the required attenuation by the passive protection device.
\item What is the maximum possible particle density that can impact the protection device according to the failure scenario? At which repetition rate can the failure happen? This will define the required robustness of the device and thus the maximum possible material density. Together with the required attenuation, this gives the required length of the device.
\item What is the aperture that has to be protected? If it has to be set a few $\sigma$ from the beam centre, it will have to be made movable. The required protection settings need to be evaluated.
\end{itemize}
The design of passive protection devices involves several different disciplines in accelerator physics and material science. Simulations for particle tracking, energy deposition in material, and thermomechanical response are required.
\subsubsection{Examples of passive protection devices in the LHC injection protection system}
\begin{figure}
\centering\includegraphics[width=.7\linewidth]{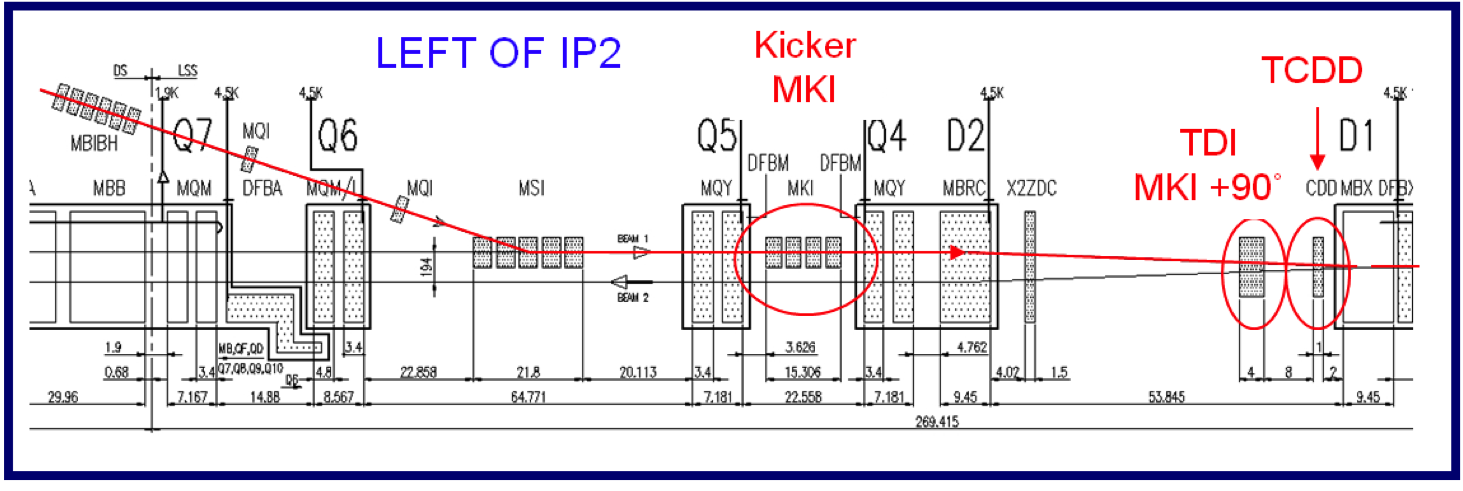}
\caption{A top view of the LHC injection region with the horizontal MSI injection septum, the vertical inject kickers MKI, followed 90$^{\circ}$ downstream by the TDI injection protection device and the TCDD mask. The red line indicates the beam direction. Part of the transfer line can also still be seen.}

\label{fig:tdi_injection}
\end{figure}

If absorbers are to protect against failures of one particular dipole error source, they are placed at 90$^{\circ}$ phase advance downstream of the error location. This is the case of the TDI absorber, which protects against LHC injection kicker failures. It has to withstand the high-intensity LHC beams and is therefore made of low-intensity materials, to be sufficiently robust. It consists of a sandwich of materials, starting with the low-density BN5000 ($\rho = 1.92\Ug/\UcmZ^{3}$) over a length of 2.85\Um, then 0.6\Um\ of \EAl, and finally 0.7\Um\ of \ECu--\EBe\ alloy. A large quantity of secondary showers still escapes the TDI if there is an impact. The TCDD   mask protects the next magnetic element, the superconducting D1 separation dipole, from the shower particles. Figure~\ref{fig:tdi_injection} shows an overview of the injection region  at LHC point 2 with the injection equipment, the TDI protection device, and the TCDD mask.

\begin{figure}
\centering\includegraphics[width=.5\linewidth]{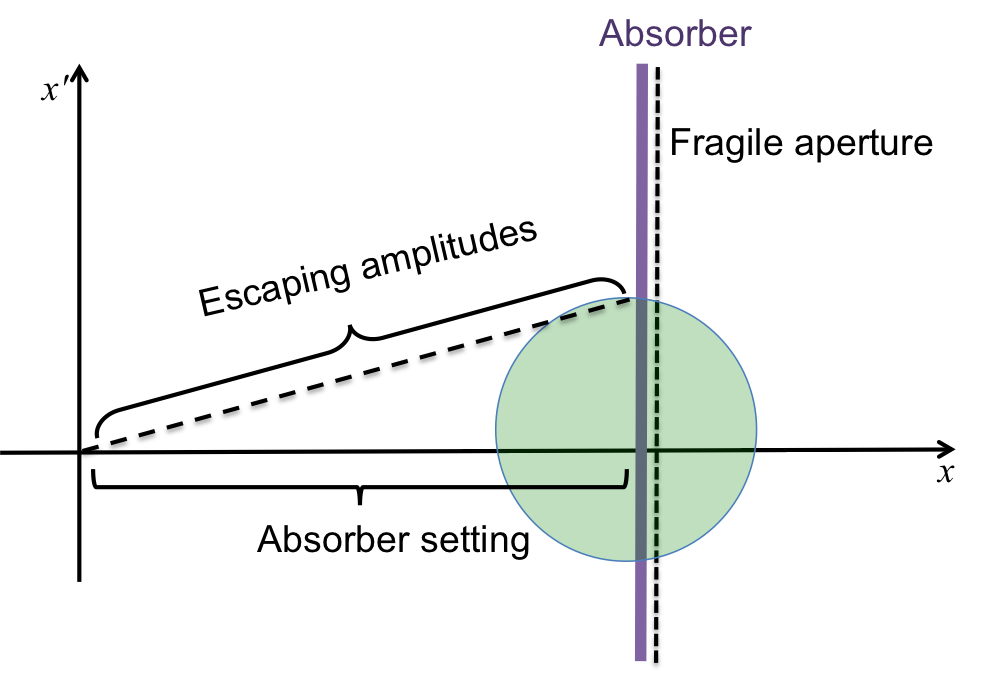}
\caption{A fraction of the surviving beam will still have larger oscillation amplitudes than the absorber setting. This has to be taken into account when choosing an appropriate absorber protection setting.}

\label{fig:tdi_setting}
\end{figure}

The setting of a protection device, \ie the distance from the nominal orbit or trajectory and thus the amplitude cut in phase-space, has to be chosen so as to include a margin for orbit variations, beta beat, mechanical tolerances, set-up tolerances, \etc The extent of the particle distribution in phase-space also has an impact on the required setting. This is illustrated in Fig. \ref{fig:tdi_setting}. If the phase-space is cut at a single location, a fraction of the surviving particles will still have oscillation amplitudes larger than the absorber setting. This effect can be reduced by adding another absorber a few tens of degrees further downstream. In the LHC injection region, the TDI absorber is complemented by two shorter graphite auxiliary collimators of 1\Um\ length at $\pm 20^{\circ}$ phase advance from the TDI, with two jaws each.

\begin{figure}
\centering\includegraphics[width=.45\linewidth]{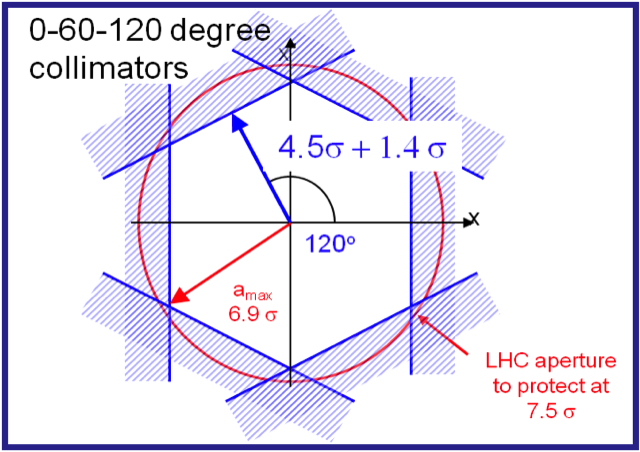}
\caption{The phase-space coverage of the LHC transfer line collimation with three collimators per plane and 60$^{\circ}$ phase advance between two collimators. The nominal setting is 4.5$\sigma$ and the set-up accuracy from collimator and trajectory in a single pass  is 1.4$\sigma$. The maximum escaping amplitudes for primary beam are, therefore, 6.9$\sigma$.}

\label{fig:tcdi}
\end{figure}

The LHC transfer lines are equipped with a generic passive protection system to protect against any failure during SPS extraction or transfer. It provides full phase-space coverage in the horizontal and vertical planes. In a single pass, this can only be achieved by installing collimators at several phase-space locations. In the case of the LHC transfer lines, three collimators per plane with 60$^{\circ}$ phase advance between two neighbouring collimators are chosen. They are located at the ends of the lines, to cover as many  failures as possible. The maximum amplitude escaping such a system for 60$^{\circ}$ between the collimators can be calculated from the collimator setting $n [\sigma]$:
\begin{equation}
a_{\mathrm{max}}\ [\sigma] = \frac{n\ [\sigma]}{\cos (60^{\circ}/2)}\ .
\end{equation}
An illustration of the phase-space cut for the LHC transfer line collimation system with a setting of 4.5$\sigma$ and 1.4$\sigma$ set-up accuracy is shown in Fig. \ref{fig:tcdi}.

\subsection{Beam dumping}
The beam dump system is one of the main components of the machine protection system. It is connected to the machine's beam permit and is triggered when the permit disappears. Fast-rising kicker magnets (and sometimes septa)  are used, as for a fast extraction system, to steer the beam onto a beam dump block. The beam dump block can be installed in the circular machine, if it is an internal beam dump, or at the end of a transfer line, if it is an external beam dump. The system has to work for all energies and beam types of the accelerator. As kicker systems are involved, the beam dump system itself can cause very fast and hence very severe failures. The design has to take the failure scenarios in account and foresee sufficient redundancy and surveillance. In the LHC, the allowable failure rate is $10^{-7}/\UhZ$ to $10^{-8}/\UhZ$, which corresponds to safety integrity level 3.

In contrast with a pure fast extraction system, where the pulse forming networks only have to be charged a few milliseconds before the extraction request, the beam dump pulse forming networks are always charged.  The charging voltage follows the energy of the accelerator. A system to derive the energy from the synchrotron is required. In the LHC and SPS, this system is called the   beam energy tracking system (BETS). The energy information is derived from the main dipole currents \cite{bib:bets}. The LHC BETS surveys the correct energy tracking of the pulse forming networks and other involved equipment, such as the dump septa currents. If a tracking error occurs, the beam dump is triggered immediately before the error becomes too large.

\begin{figure}
\centering\includegraphics[width=.9\linewidth]{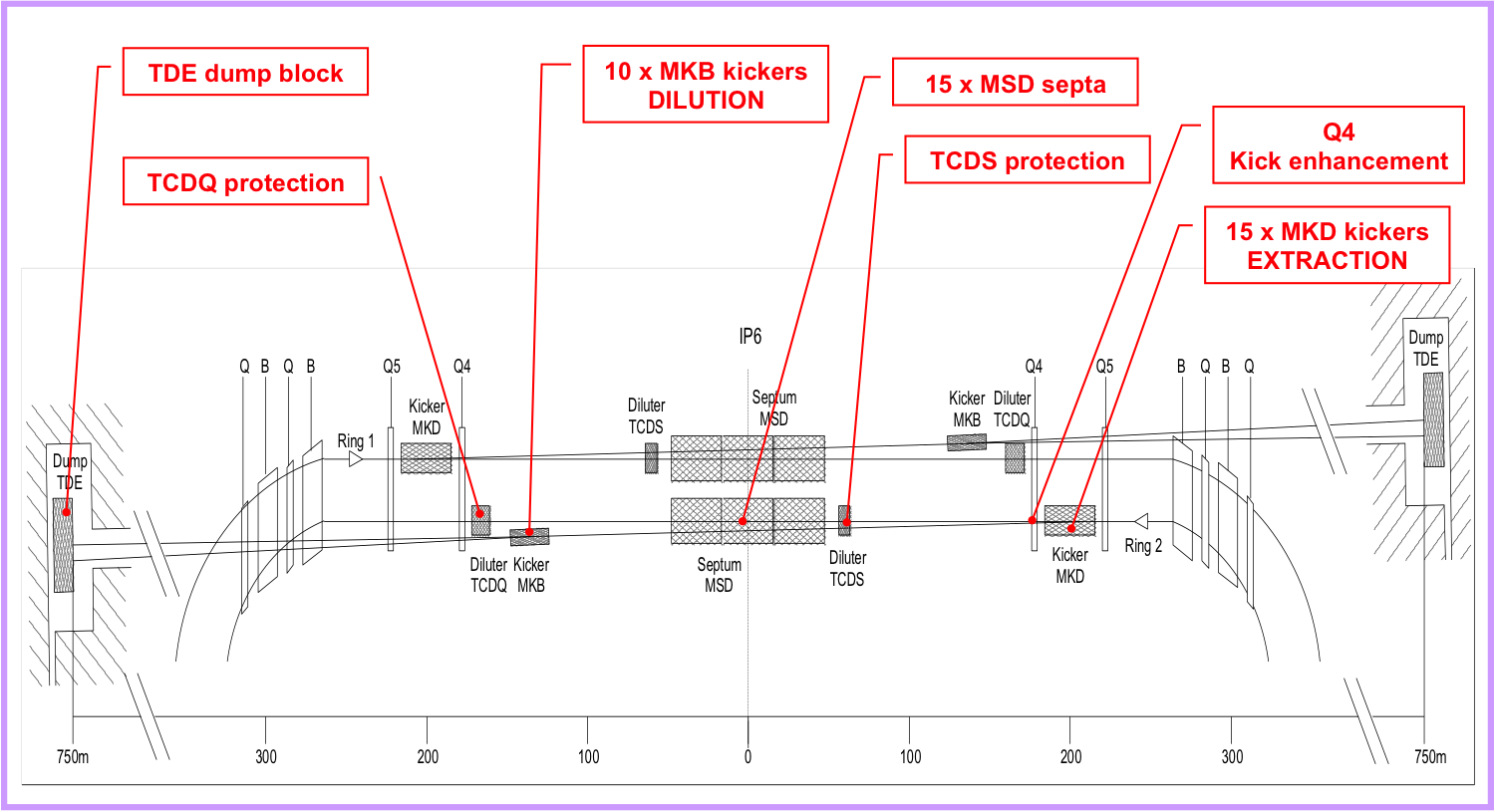}
\caption{The two beam dump systems installed in long straight section 6 of the LHC for beams 1 and 2}

\label{fig:lss6}
\end{figure}

The LHC beam dump system is installed in long straight section 6. The distance between the kicker magnets and the dump block is 975\Um. An overview of long straight section 6, the dump equipment and dump lines is shown in Fig. \ref{fig:lss6}.

The beam dump block itself is made of 0.7\Um\ and 3.5\Um\ lengths of graphite with a density of 1.73\Ug/\UcmZ$^3$, interspersed with 3.5\Um\ lengths of lower-density graphite ($\rho = 1.1\Ug/\Ucm^{3}$). This is followed by 1\Um\ of \EAl\ and 2\Um\ of \EFe\ at the end. The core is in a steel cylinder and in inert \EN$_2$ gas. It is surrounded by about \Unit{900}{t} of radiation shielding blocks. The beam dump is designed such that no structural damage is to be expected during 20\Uy\ of operation with ultimate LHC intensities ($1.7 \times 10^{11}$ protons per bunch in 3.5\Uum\ emittance and 2808 bunches). Nevertheless, the dump block is only sufficiently robust for 7\UTeV\ protons with beam dilution.

\begin{figure}
\centering\includegraphics[width=.5\linewidth]{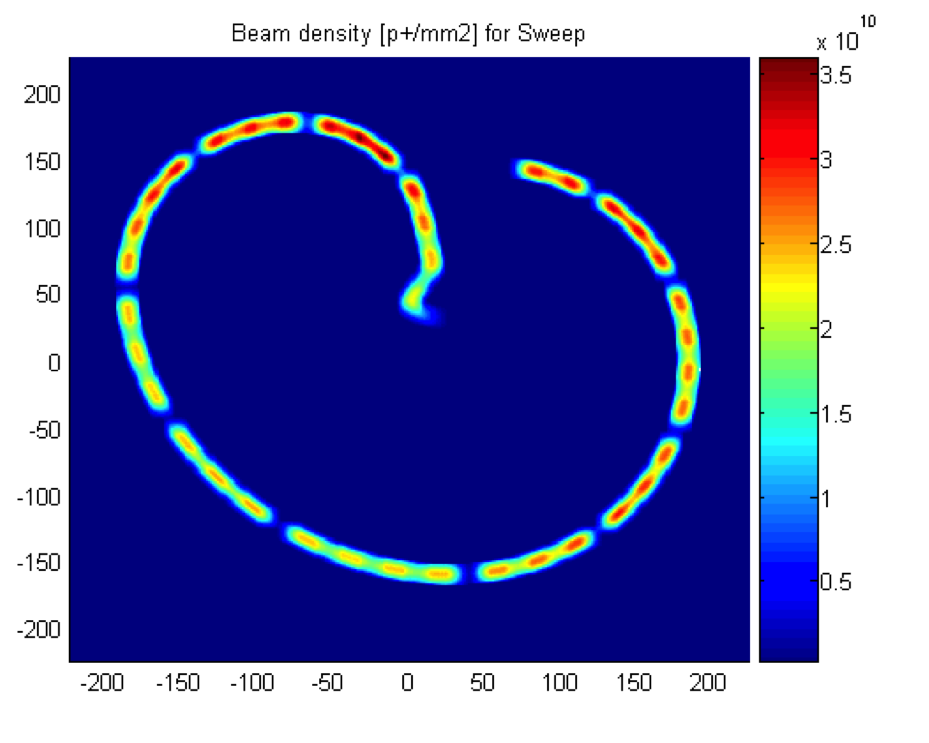}
\caption{Simulated swept beam density on the front face of the LHC beam dump block. The sweep is the result of the excitation of horizontal and vertical dilution kickers with waveforms as illustrated in Fig. \ref{fig:mkb_waveform}.}
\label{fig:mkb_sweep}
\end{figure}

\begin{figure}
\centering\includegraphics[width=.5\linewidth]{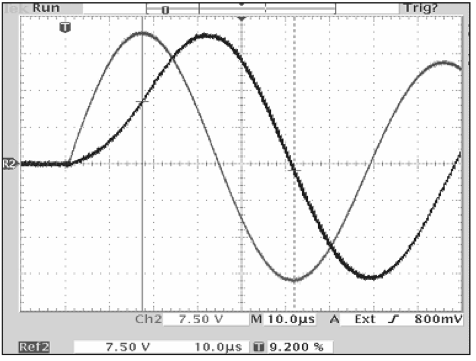}
\caption{LHC horizontal and vertical dilution kicker waveforms}
\label{fig:mkb_waveform}
\end{figure}

A set of vertical and horizontal dilution kickers is installed in the extraction channel, to sweep the beam onto the beam dump block. Not all bunches, therefore, will end up on the same spot. The resulting LHC beam dump sweep shape is shown in Fig. \ref{fig:mkb_sweep}, and the waveforms of the horizontal and vertical dilution kickers are shown in Fig. \ref{fig:mkb_waveform}.

\subsubsection{Failure scenarios of the LHC beam dump system}
\begin{figure}
\centering\includegraphics[width=.9\linewidth]{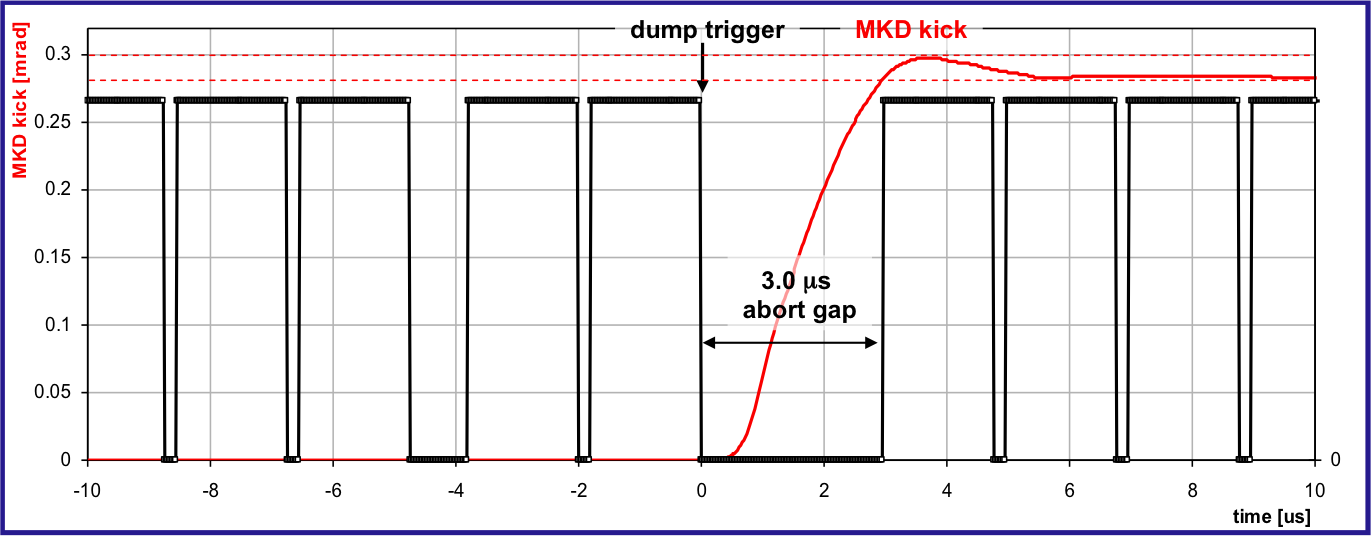}
\caption{No particles are allowed in the abort gap of the dump kickers. The black line indicates the particle bunch trains. The smallest gaps between the trains correspond to the injection kicker rise time in the SPS; the $1\Uus$ gaps come from the LHC injection kicker rise time requirements. There is one large gap of $3\Uus$ for the dump kicker rise time. The beam must not be injected into the gap. Mechanisms are arranged to prevent this and any uncaptured beam is cleaned out of the gap after transverse feedback. The red line indicates the dump kicker (MKD) waveform. }

\label{fig:abort_gap}
\end{figure}

The LHC dump kicker magnets have a rise time of 3\Uus. No particles are allowed in the so-called abort gap, whose length corresponds to the kicker rise time, Fig. \ref{fig:abort_gap}. Even if the gap is kept particle-free, the kickers can still trigger spontaneously or asynchronously. Passive protection is therefore installed at the front face of the septum magnets (TCDS) and 90$^{\circ}$ downstream of the MKD dump kickers to protect the LHC circulating aperture (TCDQ). The TCDQ is a 9\Um\ long single-sided movable absorber made of graphite. A fixed mask in front of the superconducting Q4 quadrupole protects the Q4 from damage from secondary showers generated in the TCDQ in the case of an impact.

The different failure scenarios of the LHC beam dump system have been analyzed and divided into three categories.
\subsubsubsection{LHC beam dump failure scenario: beyond design}
 This category of failures should not happen in the LHC lifetime, as they would result in severe damage to the machine. Several layers of redundancy are built into the system to cover against them:
 \begin{itemize}
 \item not receiving the trigger from the beam interlock system after a failure in the LHC;
 \item failure of beam energy tracking and extracting the entire beam at the wrong angle;
 \item fewer than 14 of the 15 dump kicker magnets firing during  a dump;
 \item fewer than 3 of the 10 dilution kickers firing during a dump.
 \end{itemize}
 \subsubsubsection{LHC beam dump scenario: active surveillance}
 The beam dump system and clean dump action depend on a number of conditions that must be monitored. If these conditions degrade, a dump is issued while the conditions are still sufficient for a clean dump. Examples are:
 \begin{itemize}
 \item failures of general services (electricity, vacuum, cooling, ethernet, \dots);
 \item bad beam position in dump region;
 \item magnet powering failure of extraction equipment. \end{itemize}
 \subsubsubsection{LHC beam dump scenario: passive protection}
 This category of failures cannot be prevented by surveillance systems and involve failures of the kicker systems.  Redundancy has been built into the number of required dump kickers to ensure correct ex\-traction. These failures can be tolerated and will not cause damage:
 \begin{itemize}
 \item one missing extraction kicker magnet;
 \item missing dilution kicker magnets;
 \item erratic behaviour of a dilution kicker magnet.
 \end{itemize}
Finally there is a set of possible kicker failures where passive protection is required. They cannot be tolerated without correct settings of the passive protection element TCDQ:
\begin{itemize}
\item erratic behaviour of an extraction kicker magnet (spurious asynchronous trigger).
\end{itemize}

\section{Final remarks}
A number of additional concepts have been invented around and for the machine protection of LHC beam transfer systems, which have not been introduced in this lecture. Some of these will be covered in the lecture by J. Wenninger on \textit{Machine Protection and Operation for the LHC}. A very important issue concerns the automatic \textit{post-operational checks} and post-mortem system, which verify the correct functioning of the relevant systems after each beam transfer by launching a quick analysis of a number of recorded datasets from equipment and beam instrumentation \cite{bib:xpoc, bib:iqc}. Degradation of the systems and increased likelihood of a failure are detected by these checks. This allows experts to intervene before an actual failure occurs.

\end{document}